%% file: main.tex
\documentclass{article}

\usepackage{microtype}
\usepackage{graphicx}
\usepackage{subfigure}
\usepackage{booktabs} 

\usepackage{hyperref}

\usepackage[accepted]{icml2024}

\usepackage{amsmath}
\usepackage{amssymb}
\usepackage{mathtools}
\usepackage{amsthm}

\usepackage[capitalize,noabbrev]{cleveref}

\theoremstyle{plain}

\theoremstyle{definition}

\theoremstyle{remark}

\usepackage[textsize=tiny]{todonotes}

\input{math_commands.tex}

\usepackage{bbm}

\newcommand{\model}{\text{MCF}}

\newcommand{\ie}{\textit{i}.\textit{e}. \ }
\newcommand{\eg}{\textit{e}.\textit{g}.\ }
\newcommand{\angstrom}{\mbox{\normalfont\AA}}

\icmltitlerunning{Swallowing the Bitter Pill: Simplified Scalable Conformer Generation}

\begin{document}

\twocolumn[
\icmltitle{Swallowing the Bitter Pill: Simplified Scalable Conformer Generation}



\icmlsetsymbol{equal}{*}

\begin{icmlauthorlist}
\icmlauthor{Yuyang Wang}{appl}
\icmlauthor{Ahmed A. Elhag}{appl,intern}
\icmlauthor{Navdeep Jaitly}{appl}
\icmlauthor{Joshua M. Susskind}{appl}
\icmlauthor{Miguel \'Angel Bautista}{appl}
\end{icmlauthorlist}

\icmlaffiliation{appl}{Apple}
\icmlaffiliation{intern}{Work was completed while A.A.E was an intern with Apple}

\icmlcorrespondingauthor{}{\{yuyang\_wang4, aa\_elhag, njaitly, jsusskind, mbautistamartin\}@apple.com}

\icmlkeywords{Machine Learning, ICML}

\vskip 0.3in
]



\printAffiliationsAndNotice{}  
\begin{abstract}
We present a novel way to predict molecular conformers through a simple formulation that sidesteps many of the heuristics of prior works and achieves state of the art results by using the advantages of scale. By training a diffusion generative model directly on 3D atomic positions without making assumptions about the explicit structure of molecules (\eg modeling torsional angles) we are able to radically simplify structure learning, and make it trivial to scale up the model sizes. This model, called Molecular Conformer Fields ({\model}), works by parameterizing conformer structures as functions that map elements from a molecular graph directly to their 3D location in space. This formulation allows us to boil down the essence of structure prediction to learning a distribution over functions. Experimental results show that scaling up the model capacity leads to large gains in generalization performance \textit{without enforcing inductive biases} like rotational equivariance. {\model} represents an advance in extending diffusion models to handle complex scientific problems in a conceptually simple, scalable and effective manner. 
\end{abstract}

\section{Introduction}
In this paper we tackle the problem of molecular conformer generation, \ie  predicting the diverse low-energy three-dimensional conformers of molecules. Molecular conformer generation is a fundamental problem in computational drug discovery and chemo-informatics, where understanding the intricate interactions between molecular and protein structures in 3D space is critical, affecting aspects such as charge distribution, potential energy, etc. \citep{compdrugdisc}. The core challenge associated with conformer generation is the vast complexity of the 3D structure space, encompassing factors such as bond lengths and torsional angles. Despite the molecular graph dictating potential 3D conformers through specific constraints, such as bond types and spatial arrangements determined by chiral centers, the conformational space experiences exponential growth with the expansion of the graph size and the number of rotatable bonds \citep{geom}. This complicates brute force and exhaustive approaches, making them virtually unfeasible for even moderately small molecules.


\begin{figure*}
    \centering
    \includegraphics[width=0.93\textwidth]{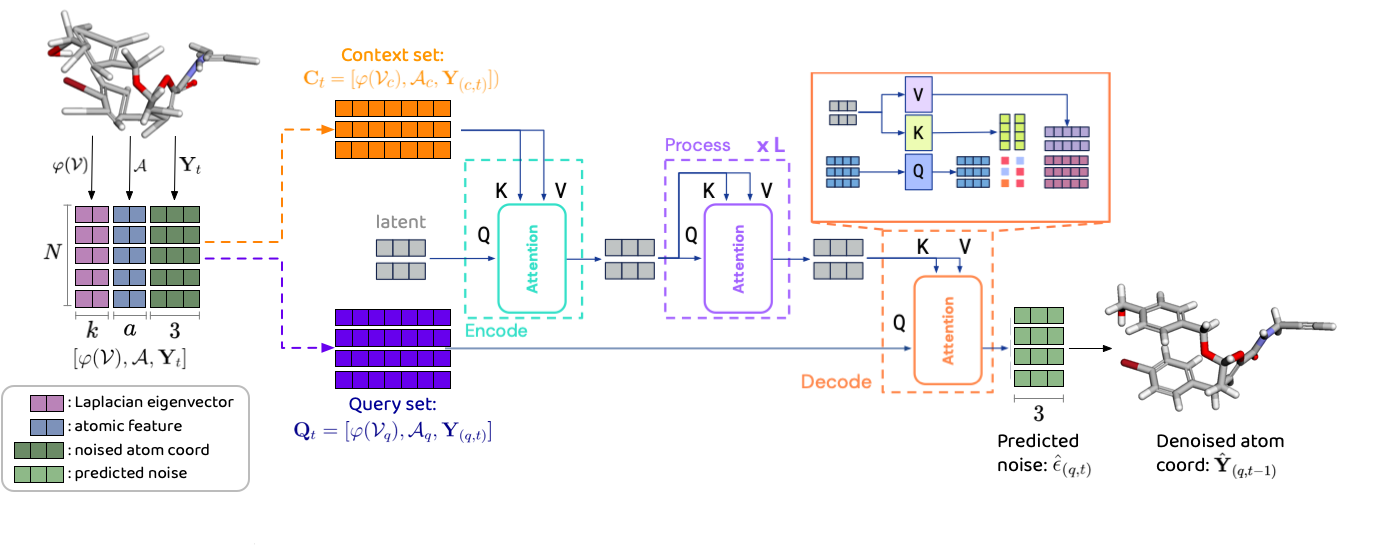}
    \caption{Overview of proposed {\model}. The structure of molecular graph is encoded through eigenvectors of Laplacian eigen-decomposition $\varphi(\mathcal{V})$ and atomic features $\mathcal{A}$. \textit{{\model} directly operates on atom coordinates in 3D space} and trains the diffusion model to denoise the function in 3D coordinates. The score network is developed with attention-based PerceiverIO architecture. Context pairs $\mathbf{C}_t$ attend to a latent array of learnable parameters via cross attention and the latent array goes through several self attention blocks. Finally, the query pairs $\mathbf{Q}_t$ cross-attend to the latent array to produce the final noise prediction $\hat{\epsilon}_q$ in 3D space. }
    \label{fig:intro}
\end{figure*}

Systematic methods, like OMEGA \citep{omega}, offer rapid processing through rule-based generators and curated torsion templates. Despite their efficiency, these models typically fail on complex molecules, as they often overlook global interactions and are tricky to extend to inputs like transition states or open-shell molecules. Classic stochastic methods, like molecular dynamics (MD) and Markov chain Monte Carlo (MCMC), rely on extensively exploring the energy landscape to find low-energy conformers. Such techniques suffer from sampling inefficiency for large molecules and struggle to generate diverse representative conformers \citep{hawkins2017conformation, mcmc, md}. In the domain of learning-based approaches, several works have looked at conformer generation problems through the lens of probabilistic modeling, using either normalizing flows \citep{cgcf} or diffusion models \citep{geodiff,torsionaldiff}. These approaches tend to use equivariant network architectures to deal with molecular graphs \citep{geodiff} or model domain-specific factors like torsional angles \citep{geomol, torsionaldiff}. However, explicitly enforcing these domain-specific inductive biases come at a cost. For example, Torsional Diffusion models rely on rule-based methods to find rotatable bonds which may fail especially for complex molecules. Ultimately, the quality of generated conformers is destined to suffer from errors of the non-differentiable cheminformatic methods used to predict local substructures. On the other hand, recent works have proposed domain-agnostic approaches for generative modeling of data in function space \citep{gem, gasp, functa, dpf} obtaining great performance. As an example, \citet{dpf} use a diffusion model to learn a distribution over functions $f$, showing great results on different data domains like images (\ie $\ f: \mathbb{R}^2 \rightarrow \mathbb{R}^3$) or 3D geometry (\ie $\ f: \mathbb{R}^3 \rightarrow \mathbb{R}^1$), where the domain of the function $\mathbb{R}^n$ is fixed across functions. Such frameworks provide a valuable paradigm to investigate whether domain-agnostic methods with little to no inductive biases can be successfully transferred to solve scientific problems (\eg molecular conformer generation). 

To this end, we present Molecular Conformer Fields ({\model}), a simple and scalable approach to learn generative models of molecular conformers.
We leverage a domain-agnostic architecture that makes no assumptions about molecular structures and trivially benefits from scale.
We formulate the molecular conformer generation problem as learning a distribution over functions/fields (we use both terms exchangeably), an approach that has been applied widely to various data domains \citep{dpf}.
Specifically, conformers are interpreted as functions that map points on graph $\mathcal{G}_i$ to atom coordinates in $\mathbb{R}^3$, $f_i: \mathcal{G}_i \rightarrow \mathbb{R}^3$, which we call a \textit{conformer field}. 
Unlike many prior efforts that shoe-horn inductive biases of molecular structures into the model (\eg developing equivariant diffusion process, modeling torsional angles, etc.) \citep{geodiff, geomol, torsionaldiff}, \textit{{\model} operates directly on 3D atom coordinates}, without enforcing molecular constraints explicitly, letting the model learn these directly from the data.

Instead of using Graph Neural Networks with intricate equivariance designs, {\model} builds a score network using PerceiverIO \citep{perceiverio} (see Fig. \ref{fig:intro}) which is a scalable and efficient variant of the Transformer architecture. Our model is simple to implement and efficient to scale. Experiments on recent conformer generation benchmarks show {\model} surpasses strong baselines by a gap that gets larger as we scale model capacity, potentially revealing a bitter lesson \cite{sutton2019bitter} moment for conformer generation, when large models with fewer domain-specific architectural inductive biases lead to better performance. Superior performance of {\model} on molecular conformation generation highlights the potential for building a singe domain-agnostic method that is simple and scalable to work on many different problems. 

Our contributions are summarized as follows: 

\begin{itemize}
\item We introduce a novel approach for molecular conformer generation that has strong scaling properties and surpasses previous methods by a large margin on standard benchmarks.
\item Our approach directly predicts the 3D position of atoms as opposed to domain-specific variables, providing a simple and scalable training recipe.
\item {\model} shows that enforcing inductive biases like rotational equivariance or modeling torsional angles is not required for generalization.




\end{itemize}

\section{Related Work}

Recent works have tackled the problem of molecular conformer generation using learning-based generative models. \citet{graphdg} and \citet{confvae} develop two-stage methods which first generate interatomic distances following VAE framework and then predict conformers based on the distances. \citet{guan2021energy} propose neural energy minimization to optimize low-quality conformers. In \citet{cgcf}, a normalizing flow approach is proposed as an alternative to VAEs. To avoid the accumulative errors from two-stage generation, \citet{confgf} implement score-based generative model to directly model the gradient of logarithm density of atomic coordinates. In GeoDiff \citep{geodiff}, a diffusion model is used which focuses on crafting equivariant forward and backward processes with equivariant graph neural networks. 
In GeoMol \citep{geomol}, the authors first predict 1-hop local structures and then propose a regression objective coupled with an Optimal Transport loss to predict the torsional angles that assemble substructures of a molecule. Following this, Torsional Diffusion \citep{torsionaldiff} proposed a diffusion model on the torsional angles of the bonds rather than a regression model used in \citet{geomol}.

Our approach extends recent efforts in generative models for functions in Euclidean space \citep{dpf, gasp, functa,gem}, to functions defined over graphs (\eg chemical structure of molecules). Different approaches have been proposed to learn distributions over fields in Euclidean space; GASP~\citep{gasp} leverages a GAN whose generator produces field data whereas a point cloud discriminator operates on discretized data and aims to differentiate real and generated functions. Two-stage approaches \citep{functa,gem} adopt a latent field parameterization \citep{deepsdf} where functions are parameterized via a hyper-network \citep{hypernetwork} and a generative model is learnt in latent space. {\model} presents a generalization over these approaches  to deal with training sets where each function $f_i$ is defined on a different graph $\mathcal{G}_i$, as opposed to in Euclidean space. In addition, {\model} also related to recent work focusing on fitting a function  on a manifold using an intrinsic coordinate system \citep{intrinsic_inr, generalized_inr}, and generalizes it to the problem of learning a probabilistic model over multiple functions defined on different graphs. Intrinsic coordinate systems have also been used in Graph Transformers to tackle supervised learning tasks \citep{lap1, lap2, lap3, lap4}.

Recent strides in the domain of protein folding dynamics have witnessed revolutionary progress, with modern methodologies capable of predicting crystallized 3D structures solely from amino-acid sequences using auto-regressive models like AlphaFold \citep{alphafold}. However, transferring these approaches seamlessly to general molecular data is fraught with challenges. Molecules present a unique set of complexities, manifested in their highly branched graphs, varying bond types, and chiral information, aspects that make the direct application of protein folding strategies to molecular data a challenging endeavor.

\section{Preliminaries}

\subsection{Diffusion Probabilistic Fields}

Diffusion Probabilistic Fields (DPF) \citep{dpf} belongs to the broad family of latent variable models \citep{latentvariablemodels} and can be consider a generalization of DDPMs \citep{ddpm} to deal with functions $f: \mathbf{M} \rightarrow Y$ which are infinite dimensional. Conceptually speaking, DPF \citep{dpf} parameterizes functions $f$ with a set of context pairs containing input-outputs to the function. Using these context pairs as input to DPF, the model is trained to denoise any query coordinate (\eg query pairs) in the domain of the function at timestep $t$ (as shown in Fig.~\ref{fig:intro}). In order to learn a parametric distribution over functions $p_\theta(f_0)$ from an empirical distribution of functions s $q(f_0)$, DPF reverses a diffusion Markov Chain that generates function latents $f_{1:T}$ by gradually adding Gaussian noise to (context) input-output pairs randomly drawn from $f \sim q(f_0)$ for $T$ time-steps as follows: $q(f_t | f_{t-1}):=\mathcal{N}\left( f_{t-1}; \sqrt{\bar{\alpha}_t}f_0, (1-\bar{\alpha}_t) \rmI \right)$. Here, $\bar{\alpha}_t$ is the cumulative product of fixed variances $\alpha_t$ with a handcrafted scheduling up to time-step $t$. DPF \citep{dpf} follows the training recipe in \citet{ddpm} in which: i) The forward process adopts sampling in closed form. ii) reversing the diffusion process is equivalent to learning a sequence of denoising (or score) networks $\epsilon_\theta$, with tied weights. Reparameterizing the forward process as $f_t = \sqrt{\bar{\alpha}_t} f_0  + \sqrt{1 - \bar{\alpha}_t}\epsilon$ results in the  ``simple'' DDPM loss: $\mathbb{E}_{t \sim [0, T], f_0 \sim q(f_0), \epsilon \sim \mathcal{N}(0, \mathbf{I})} \left[\| \epsilon - \epsilon_{\theta}( \sqrt{\bar{\alpha}_t} f_0  + \sqrt{1 - \bar{\alpha}_t}\epsilon , t) \|^2 \right]$, which makes learning of the data distribution $p_\theta(f_0)$ both efficient and scalable. At inference time, DPF computes $f_{0} \sim p_\theta(f_0)$ via ancestral sampling \citep{dpf}. Concretely, DPF starts by sampling dense query coordinates and assigning a gaussian value to them $f_T \sim \mathcal{N}(\vzero, \mathbf{I})$. Then, it iteratively applies the score network $\epsilon_{\theta}$ to denoise $f_T$, thus reversing the diffusion Markov Chain to obtain $f_0$. In practice, DPFs have obtained amazing results for signals living in an Euclidean geometry. 

\subsection{Conformers as Functions on Graphs}

Following the setting in previous work \citep{geodiff, geomol, torsionaldiff} a molecule with $n$ atoms is represented as an undirected graph $\mathcal{G} = \langle \mathcal{V} , \mathcal{E} \rangle$, where $\mathcal{V} = \{v_i\}^{n}_{i=1}$ is the set of vertices representing atoms and $\mathcal{E} = \{e_{ij} | (i, j) \subseteq |\mathcal{V}| \times |\mathcal{V}|\}$ is the set of edges representing inter-atomic bonds. We further use $\mathcal{A}$ to denote atomic features which also are leveraged by our generative model. In this paper, we parameterize a molecule's conformer as a function $f: \mathcal{G} \rightarrow \mathbb{R}^3$ that takes atoms (\eg vertices) in the molecular graph $\mathcal{G}$ and maps them to 3D space, we call this function a \textit{conformer field}. The training set is composed of conformer fields $f_i : \mathcal{G}_i \rightarrow \mathbb{R}^3$, where each field maps atoms of a different molecule $\mathcal{G}_i$ to a 3D point. We then formulate the task of conformer generation as learning a prior over a training set of conformer fields. We drop the subscript $i$ in the remainder of the text for notation simplicity.

We learn a denoising diffusion generative model \citep{ddpm} over conformer fields $f$. In particular, given conformer field samples $f_0 \sim q(f_0)$ the forward process takes the form of a Markov Chain with progressively increasing Gaussian noise: $q(f_{1:T}|f_0)=\prod_{t=1}^Tq(f_t|f_{t-1}), \quad q(f_t | f_{t-1}):=\mathcal{N}\left( f_{t-1}; \sqrt{\bar{\alpha}_t}f_0, (1-\bar{\alpha}_t) \rmI \right)$. We train {\model} using the denoising objective function in \citep{ddpm}: $\mathbb{E}_{t \sim [0, T], f_0 \sim q(f_0), \epsilon \sim \mathcal{N}(0, \mathbf{I})} \left[\| \epsilon - \epsilon_{\theta}( \sqrt{\bar{\alpha}_t} f_0  + \sqrt{1 - \bar{\alpha}_t}\epsilon , t) \|^2 \right]$.

\subsection{Equivariance in Conformer Generation}

Equivariance has become an important topic of study in generative models \citep{equiv_gen1, equiv_gen2, equiv_gen3}. In particular, enforcing equivariance as an explicit inductive bias in neural networks can lead to improved generalization \citep{equivariantlikelihood} by constraining the space of functions that can be represented by a model. On the other hand, recent literature shows that models that can learn these symmetries from data rather than explicitly enforcing them (\eg Transformers vs CNNs) tend to perform better as they are more amenable to optimization \citep{bai2021transformers}.  

Equivariance also plays an interesting role in conformer generation. On one hand, it is important when training likelihood models of conformers, as the likelihood of a conformer is invariant to roto-translations \citep{equivariantlikelihood}. On the other hand, when training models to generate conformers given a molecular graph, explicitly baking roto-translation equivariance might not be as necessary. This is because the intrinsic structure of the conformer encodes far more information about its properties than the extrinsic coordinate system (eg. rotation and translation) in which the conformer is generated \citep{qm9_1}. In addition, recent approaches for learning simulations on graphs \citep{non_equiv1} or pre-training models for molecular prediction tasks \citep{non_equiv2} have successfully relied on non-equivariant architectures.

In this paper, we ask whether inductive biases like rotational equivariance can be traded for model scale in general purposes architectures like Transformers. Our empirical results show that explicitly enforcing roto-translation equivariance is not a strong requirement for generalization. Furthermore, we show that scalable approaches that do not explicitly enforce roto-translation equivariance (like ours) can outperform approaches that do by a large margin .

\section{Method}
\label{sect:method}

{\model} is a diffusion generative model that captures distributions over conformer fields. We are given observations in the form of an empirical distribution $f_0 \sim q(f_0)$ over fields where a field $f_0: \mathcal{G} \rightarrow \mathbb{R}^3$ maps vertices $v \in \mathcal{G}$ on a molecular graph $\mathcal{G}$ to 3D space $\mathbb{R}^3$. 

\begin{figure*}[t]
\begin{minipage}[t]{0.495\textwidth}
\begin{algorithm}[H]
  \caption{Training} \label{alg:training}
  \small
  \begin{algorithmic}[1]
  \STATE  $\Delta_\mathcal{G} \varphi_i = \varphi_i \lambda_i$ // Compute Laplacian eigenvectors
    \REPEAT
      \STATE $(\rmC_0, \rmQ_0) \sim \mathrm{Uniform}(q(f_0))$
      \STATE $t \sim \mathrm{Uniform}(\{1, \dotsc, T\})$
      \STATE $\mathbf{\epsilon}_c \sim\mathcal{N}(\vzero,\mathbf{I})$, $\mathbf{\epsilon}_q \sim\mathcal{N}(\vzero,\mathbf{I})$
      \STATE $\rmC_{t} = [\varphi(\mathcal{V}_{c}), \sqrt{\bar{\alpha}_t}\rmY_{(c,0)} + \sqrt{1-\bar{\alpha}_t} \mathbf{\epsilon}_c]$
      \STATE $\rmQ_{t} = [\varphi(\mathcal{V}_{q}), \sqrt{\bar{\alpha}_t}\rmY_{(q,0)} + \sqrt{1-\bar{\alpha}_t} \mathbf{\epsilon}_q]$
      \STATE Take gradient descent step on
      \STATE $\nabla_\theta \left\| \mathbf{\epsilon}_{q} - \epsilon_\theta(\rmC_t, t, \rmQ_t) \right\|^2$
    \UNTIL{converged}
  \end{algorithmic}
\end{algorithm}
\end{minipage}
\hfill
\begin{minipage}[t]{0.40\textwidth}
\begin{figure}[H]
    \centering
    \includegraphics[width=\textwidth]{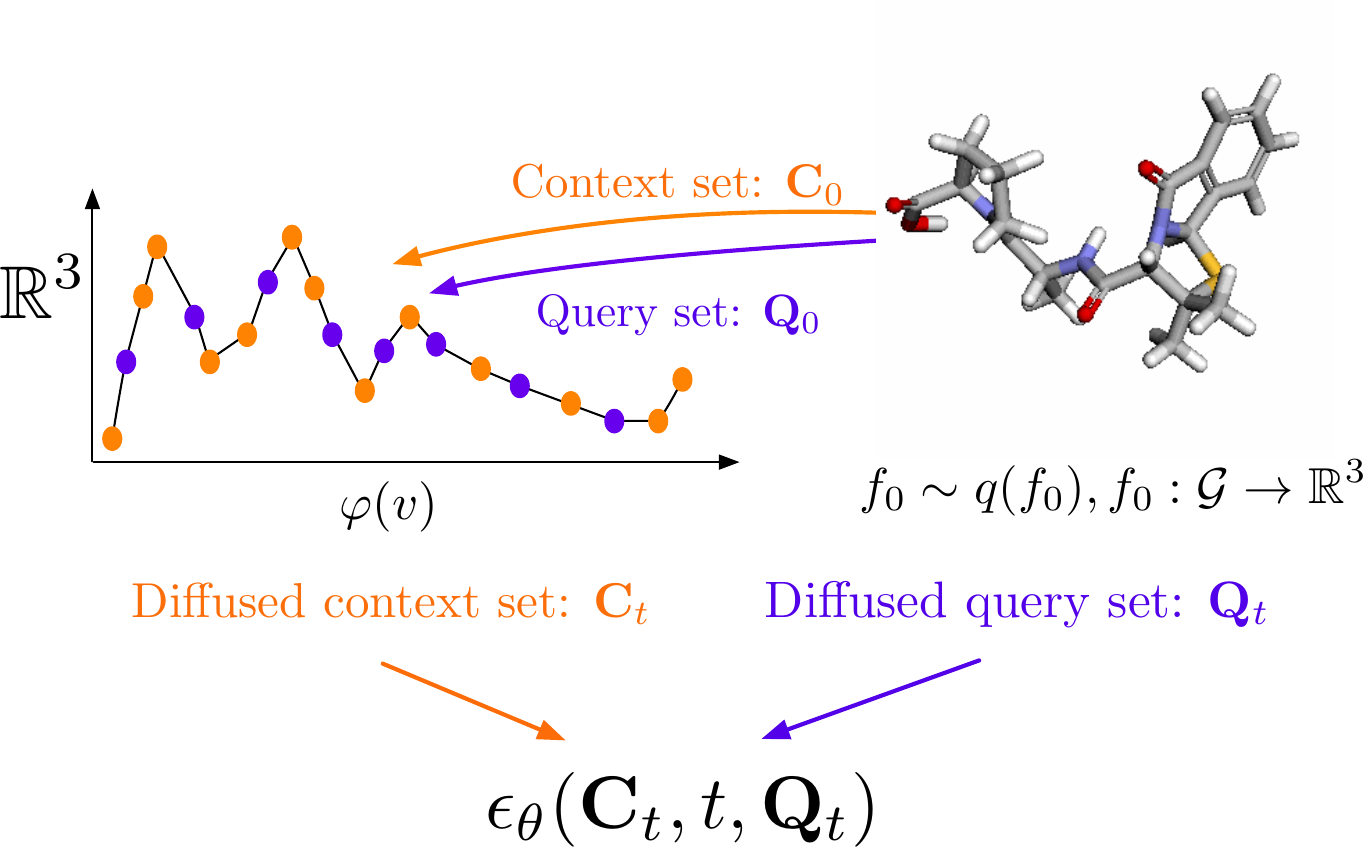}
    \label{fig:dpf_training}
\end{figure}
\end{minipage}
\caption{\textbf{Left:} {\model} training algorithm. \textbf{Right}: Visual depiction of a training iteration for a conformer field. See Sect.~\ref{sect:method} for definitions (.}
\end{figure*}

\begin{figure*}[t]
\begin{minipage}[t]{0.495\textwidth}
\begin{algorithm}[H]
  \caption{Sampling} \label{alg:sampling}
  \small
  \begin{algorithmic}[1]  
    \STATE  $\Delta_\mathcal{G} \varphi_i = \varphi_i \lambda_i$ // LBO eigen-decomposition
    \STATE $\rmQ_T = [\varphi(\mathcal{V}_q) , \rmY_{(q, t)} \sim \mathcal{N}(\vzero_q, \rmI_q)]$ 
    \STATE $\rmC_T \subseteq \rmQ_T$ \COMMENT{Random subset}
    \FOR{$t=T, \dotsc, 1$}
      \STATE $\vz \sim \mathcal{N}(\vzero, \rmI)$ if $t > 1$, else $\vz = \vzero$
      \tiny
      \STATE $\rmY_{(q, t-1)}~=  \frac{1}{\sqrt{\alpha_t}}\left( \rmY_{(q,t)} - \frac{1-\alpha_t}{\sqrt{1-\bar{\alpha}_t}} \epsilon_\theta(\rmC_t, t, \rmQ_t) \right) + \sigma_t \vz$
      \small
      \STATE $\rmQ_{t-1} = [\varphi(\mathcal{V}_q), \rmY_{(q, t-1)}]$
     \STATE $\rmC_{t-1} \subseteq \rmQ_{t-1}$ \COMMENT{Same subset as in step 2}
      \small
    \ENDFOR
    \STATE \textbf{return} $f_0$ evaluated at coordinates $\varphi(\mathcal{V}_q)$
    \vspace{.04in}
  \end{algorithmic}
\end{algorithm}
\end{minipage}
\hfill
\begin{minipage}[t]{0.40\textwidth}
\begin{figure}[H]
    \includegraphics[width=\textwidth]{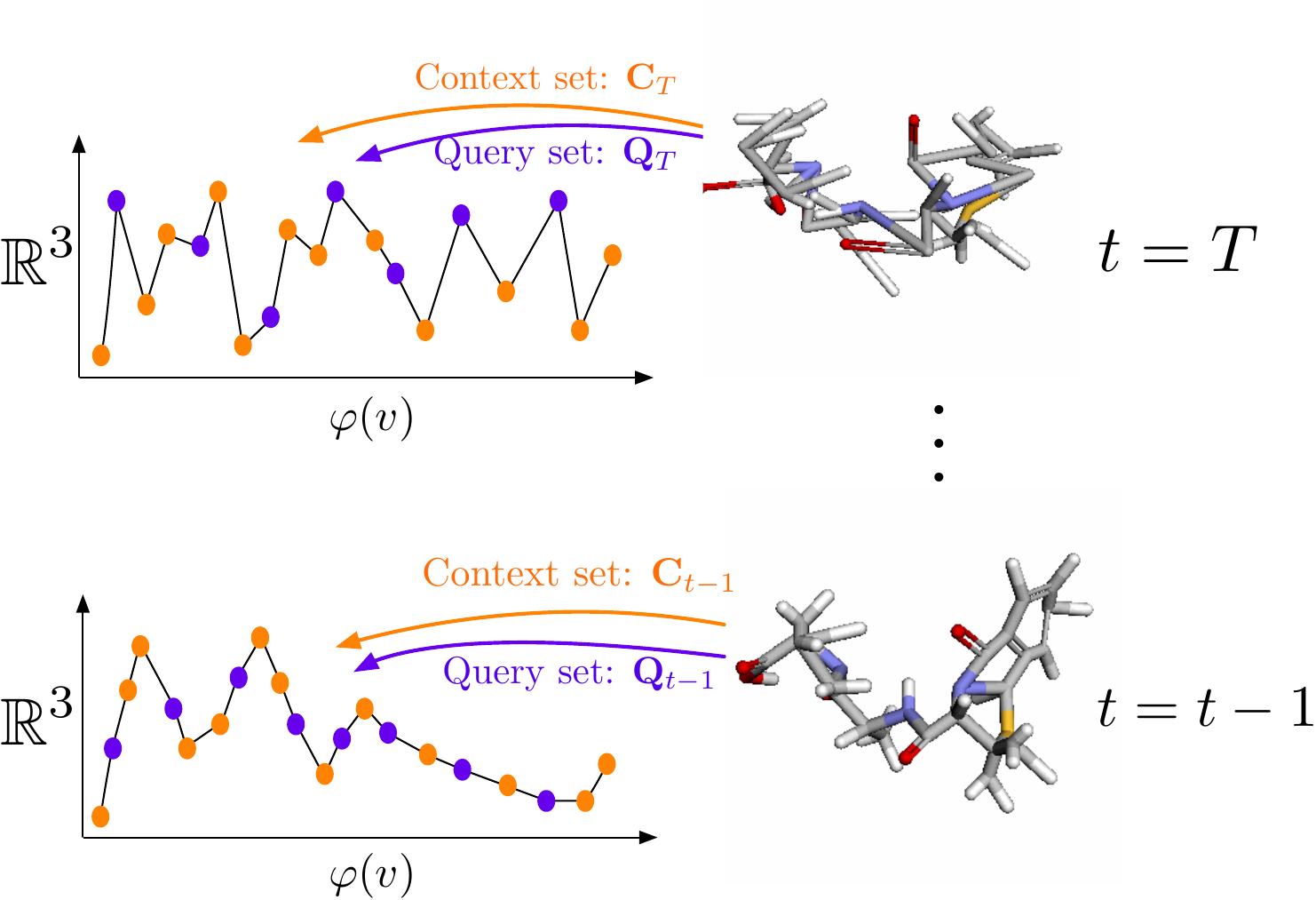}
    \label{fig:dpf_inference}
\end{figure}
\end{minipage}
\caption{\textbf{Left:} {\model} sampling algorithm. \textbf{Right}: Visual depiction of the sampling process of a conformer field.}

\end{figure*}

To tackle the problem of learning a diffusion generative model over conformer fields we extend the recipe in DPF \citep{dpf}, generalizing from fields defined in ambient Euclidean space to functions on graphs (\eg conformer fields). In order to do this, we compute the $k$ leading eigenvectors of the normalized graph Laplacian $\Delta_\mathcal{G}$ \citep{lap1, lap2} as positional encoding for points in the graph. The eigen-decomposition of the normalized graph Laplacian can be computed efficiently using sparse eigen-problem solvers \citep{sparse_eigen} and only needs to be computed once before training. We use the term $\varphi(v) = \sqrt{n}[\varphi_1(v), \varphi_2(v), \dots, \varphi_k(v)] \in \mathbb{R}^k$ to denote the normalized Laplacian eigenvector representation of a vertex $v \in \mathcal{G}$. 

We adopt an explicit field parametrization where a field is characterized by uniformly sampling a set of vertex-signal pairs $\{(\varphi(v_c), \vy_{(c,0)})\}$, $v_c \in \mathcal{G}, \vy_{(c, 0)} \in \mathbb{R}^3$, which is denoted as \textit{context set}. We row-wise stack the context set and refer to the resulting matrix via $\rmC_0~=~[\varphi(\mathcal{V}_c),~\rmY_{(c,0)}]$. Here, $\varphi(\mathcal{V}_c)$ denotes the Laplacian eigenvector representation context vertices and $\rmY_{(c,0)}$ denotes the 3D position of context vertices at time $t=0$. We define the forward process for the context set by diffusing the 3D positions and keeping Laplacian eigenvectors fixed:

\begin{equation}
\label{eq:forward_pairs}
\rmC_{t} = [\varphi(\mathcal{V}_{c}), \rmY_{(c,t)}=\sqrt{\bar{\alpha}_t}\rmY_{(c,0)} + \sqrt{1-\bar{\alpha}_t} \mathbf{\epsilon}_c],
\end{equation}

where $\mathbf{\epsilon}_c \sim \mathcal{N}(\vzero, \mathbf{I})$ is a noise vector of the appropriate size. We now turn to the task of formulating a score network for fields. The score network needs to take as input the context set (\ie \ the field parametrization), and needs to accept being evaluated for any point in $\mathcal{G}$. We do this by sampling a \textit{query set} of vertex-signal pairs $\{ \varphi(v_q), \vy_{(q,0)}\}$. Equivalently to the context set, we row-wise stack query pairs and denote the resulting matrix as $\rmQ_0~=~[\varphi(\mathcal{V}_q),~\rmY_{(q,0)}]$. Note that the forward diffusion process is equivalently defined for both context and query sets: 

\begin{equation}
\label{eq:forward_query}
\rmQ_{t} = [\varphi(\mathcal{V}_{q}), \rmY_{(q,t)}=\sqrt{\bar{\alpha}_t}\rmY_{(q,0)} + \sqrt{1-\bar{\alpha}_t} \mathbf{\epsilon}_q],
\end{equation}
 
where $\mathbf{\epsilon}_q \sim \mathcal{N}(\vzero, \mathbf{I})$ is a noise vector of the appropriate size. The underlying field is solely defined by the context set, and the query set are the function evaluations to be de-noised. The resulting \textit{score field} model is formulated as follows, $\hat{\mathbf{\epsilon}_q} = \epsilon_\theta(\rmC_t, t, \rmQ_t)$.

Using the explicit field characterization and the score field network, we obtain the training and inference procedures in Alg.~\ref{alg:training} and Alg.~\ref{alg:sampling}, respectively, which are accompanied by illustrative examples of sampling a conformer field. For training, we uniformly sample context and query sets from $f_0 \sim \mathrm{Uniform}(q(f_0))$ and only corrupt their signal using the forward process in Eq.~\eqref{eq:forward_pairs} and Eq.~\eqref{eq:forward_query}. We train the score field network $\epsilon_\theta$ to denoise the signal portion of the query set, given the context set. During sampling, to generate a conformer fields $f_0 \sim p_\theta(f_0)$ we first define a query set $\rmQ_T~=~[\varphi(\mathcal{V}_q),~\rmY_{(q, T)} \sim~\mathcal{N}(\vzero,~\rmI)]$ of random atom positions to be de-noised. We set the context set to be a random subset of the query set. We use the context set to denoise the query set and follow ancestral sampling as in the vanilla DDPM \citep{ddpm}. Note that during inference the eigen-function representation $\varphi(v)$ of the context and query sets does not change, only their corresponding signal value (\eg \ their 3D position).

\subsection{Score Field Network $\epsilon_\theta$}

In {\model}, the score field's design space covers all architectures that can process irregularly sampled data, such as Transformers \citep{transformers} and their corresponding Graph counterparts \citep{lap1, lap2, lap3, lap4} which have recently gained popularity in the supervised learning setting. The score field network $\epsilon_\theta$ is primarily implemented using PerceiverIO~\citep{perceiverio}, an effective Transformer encoder-decoder architecture. A PerceiverIO is chosen due to its nature of a general-purposed architecture that can handle data of a wide variety domains. It provides a suitable test bed for evaluating how well models without domain-specific inductive bias (\eg equivariance) perform in solving scientific problems (\eg molecular conformer generation as investigated in this work). PerceiverIO encodes interactions between elements in sets using attention, which has been demonstrated to be scalable in many previous works \citep{gpt3}. Fig.~\ref{fig:intro} demonstrates how these sets are used within the PerceiverIO architecture. To elaborate, the encoder maps the context set into latent arrays (\ie a group of learnable vectors) through a cross-attention layer, while the decoder does the same for query set. For a more detailed analysis of the PerceiverIO architecture refer to~\cite{perceiverio}.The time-step $t$ is incorporated into the score computation by concatenating a positional embedding representation of $t$ to both context and query sets. 

\section{Experiments}

We use two popular datasets: GEOM-QM9 and GEOM-DRUGS \citep{geom}. Datasets are preprocessed and split as described in \citet{geomol}. We deploy PerceiverIO with small (S), base (B) and large (L) sizes, which contain 13M, 64M and 242M parameters respectively. More implementation details can be found in Appendix \ref{app:implementation_details}. We provide additional experiments that validate the design choices for the score network architecture, as well as empirically validating the chemical properties of generated conformers in the Appendix \ref{app:additional_experiments}.

\subsection{GEOM-QM9}
\label{sect:qm9}

\begin{table}[t]
\setlength{\tabcolsep}{3pt}
    \centering
    \scriptsize
\begin{tabular}{l cc cc cc cc}
& \multicolumn{4}{c}{Recall} & \multicolumn{4}{c}{Precision} \\   
\toprule
& \multicolumn{2}{c}{COV $\uparrow$} & \multicolumn{2}{c}{AMR $\downarrow$}  & \multicolumn{2}{c}{COV $\uparrow$} & \multicolumn{2}{c}{AMR $\downarrow$} \\
\toprule
& mean & median & mean & median & mean & median & mean & median \\
\midrule
CGCF & 69.5 & 96.2 & 0.425 & 0.374 & 38.2 & 33.3 & 0.711 & 0.695 \\
GeoDiff & 76.5 & \textbf{100.0} & 0.297 & 0.229 & 50.0 & 33.5 & 0.524 & 0.510 \\
GeoMol & 91.5 & \textbf{100.0} & 0.225 & 0.193 & 87.6 & \textbf{100.0} & 0.270 & 0.241 \\
Tor. Diff. & 92.8 & \textbf{100.0} & 0.178 & 0.147 & 92.7 & \textbf{100.0} & 0.221 & 0.195 \\
\midrule
{\model} & \textbf{95.0} & \textbf{100.0} & \textbf{0.103} & \textbf{0.044} & \textbf{93.7} & \textbf{100.0} & \textbf{0.119} & \textbf{0.055} \\
\bottomrule
\end{tabular}
\caption{Molecule conformer generation results on GEOM-QM9. {\model} obtains better results than the state-of-the-art baselines.
}
\label{tab:qm9}
\end{table}

Following the standard setting for molecule conformer prediction we use the GEOM-QM9 dataset which contains $\sim130$K molecules ranging from 3 to 29 atoms. We report our results with base size model (\ie {\model}-B) in Tab. \ref{tab:qm9} and compare with CGCF \citep{cgcf}, GeoDiff \citep{geodiff}, GeoMol \citep{geomol} and Torsional Diff. \citep{torsionaldiff}. 
Note that all baselines make strong assumptions about the geometric structure of molecules. They either develop equivariant diffusion process \citep{geodiff} or model domain-specific characteristics like interatomic distances \citep{cgcf} and torsional angles of rotatable bonds \citep{geomol, torsionaldiff}. In contrast, {\model} simply models the distribution of 3D coordinates of atoms without making any assumptions about the underlying structure. Finally we report the same metrics as Torsional Diff. \citep{torsionaldiff} to compare the generated and ground truth conformer ensembles: average minimum RMSD (AMR) and coverage (COV). These metrics are reported both for precision, measuring the accuracy of the generated conformers, and recall, measuring how well the generated ensemble covers the ground-truth ensemble (details about metrics can be found in Appendix~\ref{app:metrics}). We generate $2K$ conformers for a molecule with $K$ ground truth conformers.

Tab. \ref{tab:qm9} shows that {\model} outperforms previous approaches by a substantial margin. In addition, it is important to note that {\model} is a general approach for learning functions on graphs that does not make any assumptions about the intrinsic geometric factors important in conformers like torsional angles. This makes {\model} simpler to implement and applicable to other settings in which intrinsic geometric factors are not known or expensive to compute.

\subsection{GEOM-DRUGS}
\label{sect:drugs}

\begin{table}[t]
\setlength{\tabcolsep}{3pt}
    \centering
    \scriptsize
\begin{tabular}{l cc cc cc cc}
& \multicolumn{4}{c}{Recall} & \multicolumn{4}{c}{Precision} \\   
\toprule
& \multicolumn{2}{c}{COV $\uparrow$} & \multicolumn{2}{c}{AMR $\downarrow$}  & \multicolumn{2}{c}{COV $\uparrow$} & \multicolumn{2}{c}{AMR $\downarrow$} \\
\toprule
& mean & median & mean & median & mean & median & mean & median \\
\midrule
GeoDiff & 42.1 & 37.8 & 0.835 & 0.809 & 24.9 & 14.5 & 1.136 & 1.090 \\
GeoMol & 44.6 & 41.4 & 0.875 & 0.834 & 43.0 & 36.4 & 0.928 & 0.841 \\
Tor. Diff. & 72.7 & 80.0 & 0.582 & 0.565 & 55.2 & 56.9 & 0.778 & 0.729 \\
\midrule
{\model}-S & 79.4 & 87.5 & 0.512 & 0.492 & 57.4 & 57.6 & 0.761 & 0.715 \\
{\model}-B & 84.0 & 91.5 & 0.427 & 0.402 & 64.0 & 66.2 & 0.667 & 0.605 \\
{\model}-L & \textbf{84.7} & \textbf{92.2} & \textbf{0.390} & \textbf{0.247} & \textbf{66.8} & \textbf{71.3} & \textbf{0.618} & \textbf{0.530} \\
\bottomrule
\end{tabular}
\caption{Molecule conformer generation results on GEOM-DRUGS. {\model} surpasses state-of-the-art baselines by large margin.
}
\label{tab:drugs}
\end{table}

\begin{figure*}[t]
\centering
\setlength{\tabcolsep}{1pt}
    \begin{tabular}{ccc}
    \centering
    \includegraphics[width=0.3\textwidth]{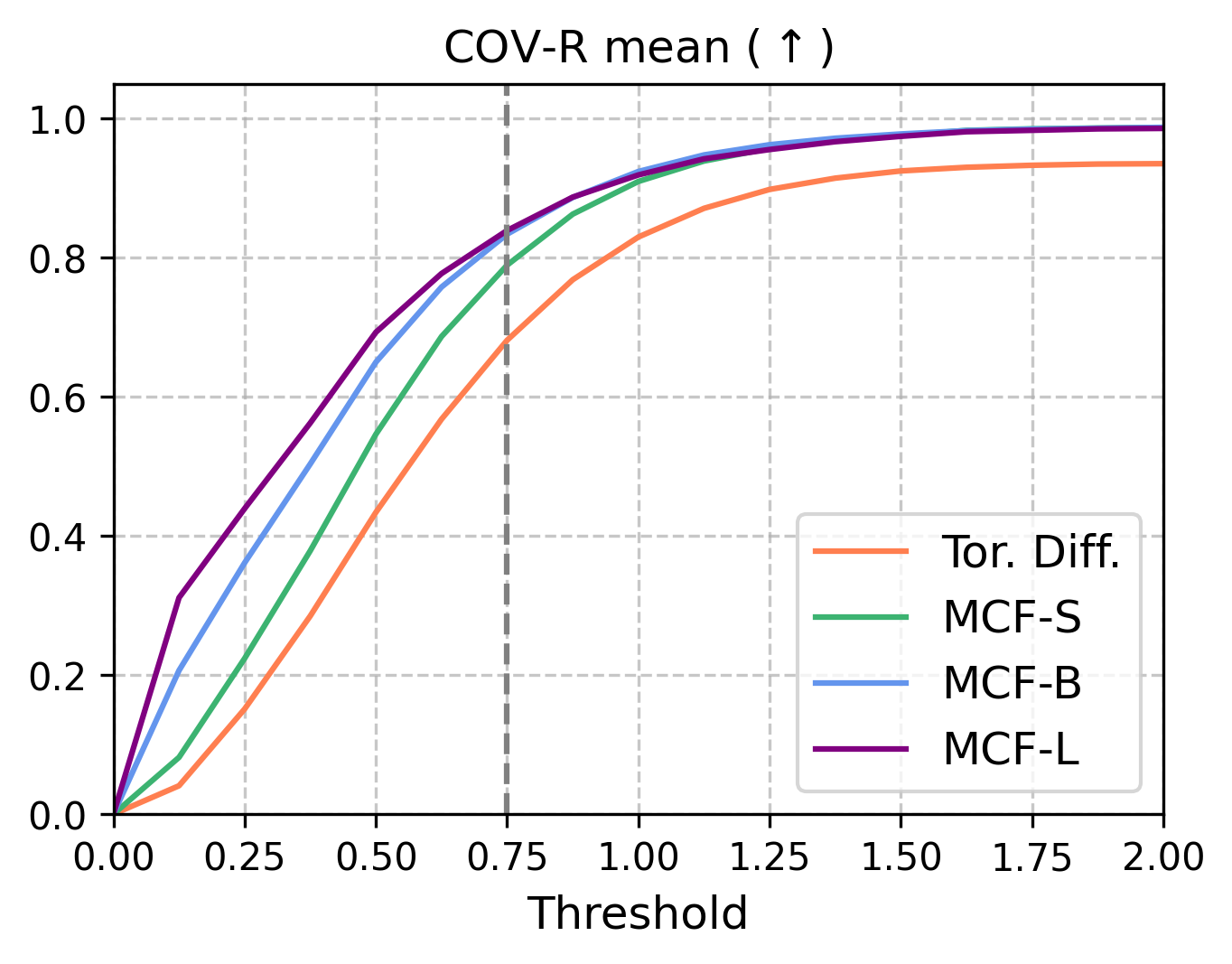} & \includegraphics[width=0.3\textwidth]{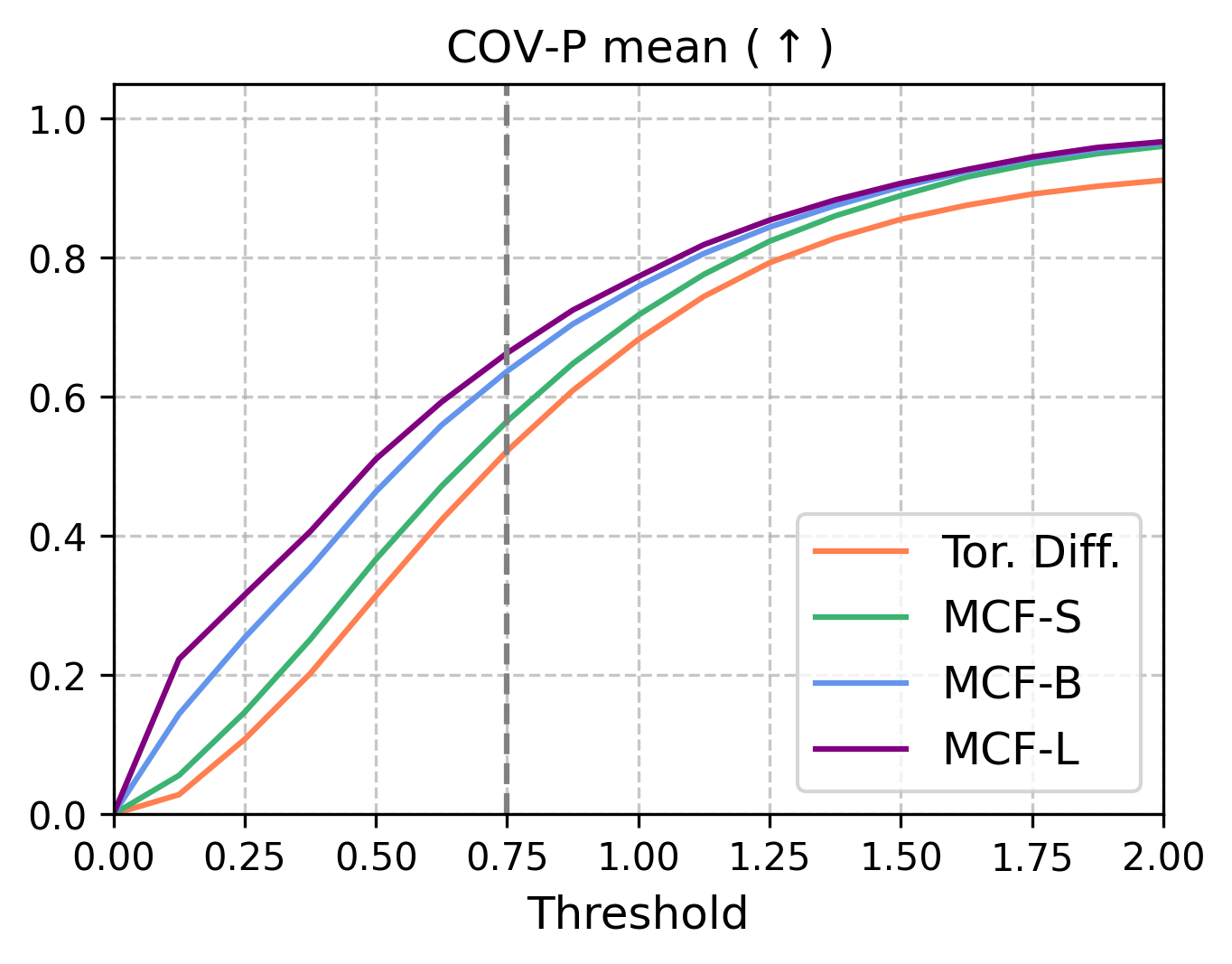} &
    \includegraphics[width=0.3\textwidth]{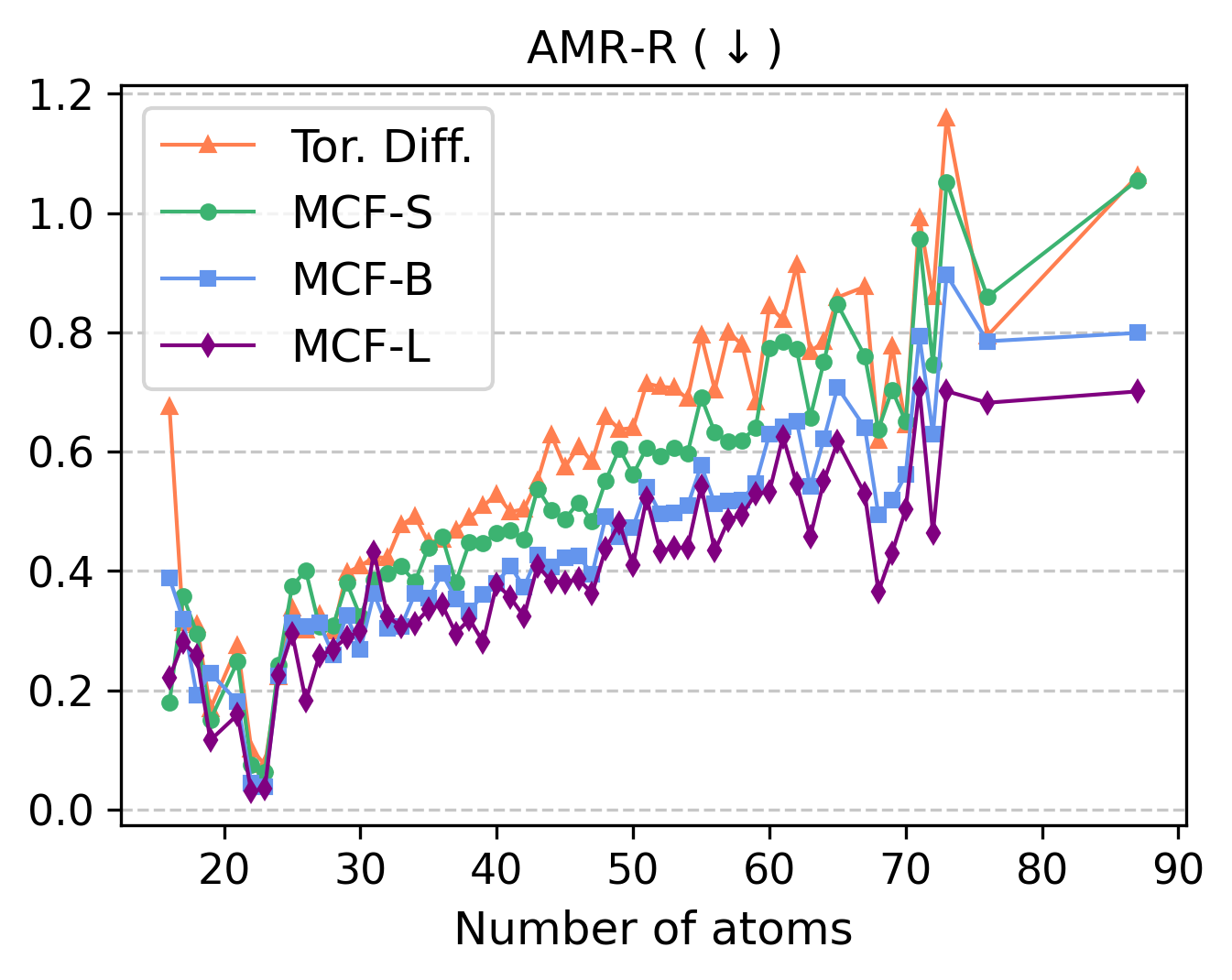} 
    \\
    \includegraphics[width=0.3\textwidth]{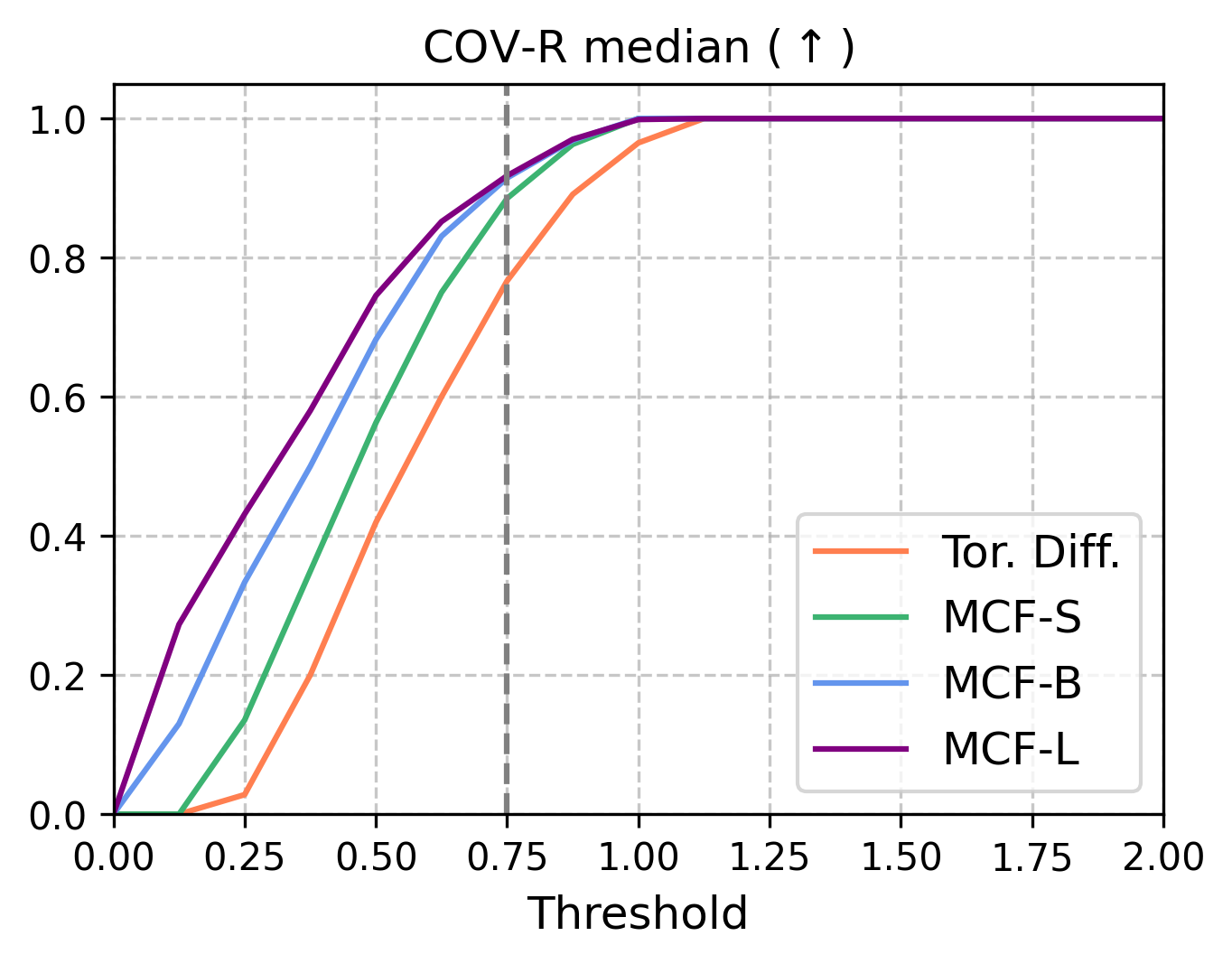} & \includegraphics[width=0.3\textwidth]{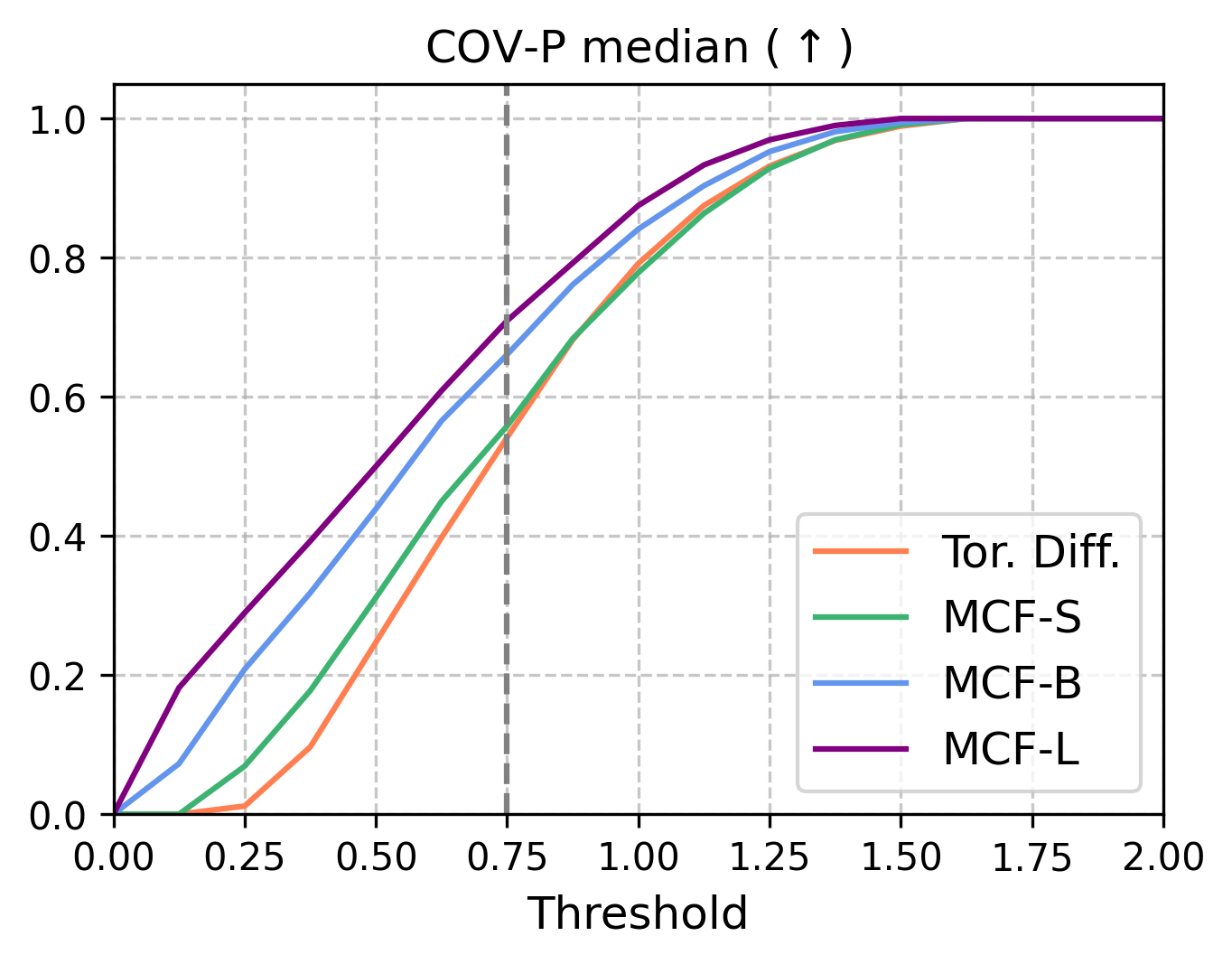} &
    \includegraphics[width=0.3\textwidth]{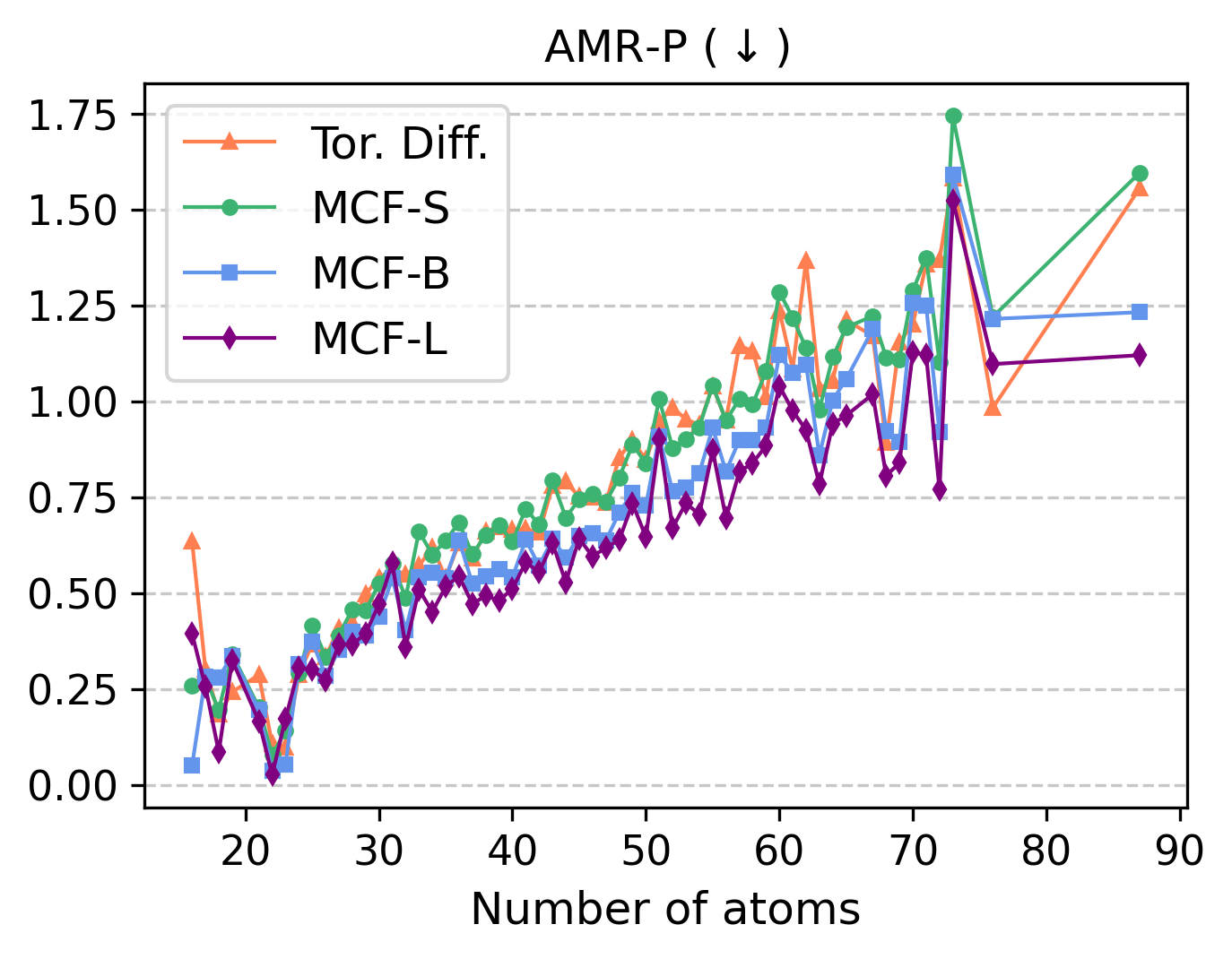} \\
        (a) & (b) & (c)
    \end{tabular}
    \caption{(a) Recall coverage and (b) precision coverage as a function of the threshold distance. {\model} outperforms Torsional Diff. across the full spectrum of thresholds. (c) Averaged AMR of recall and precision as a function of the number of atoms in molecules. }
    \label{fig:cov_func_thres}
\end{figure*}

To test the capacity of {\model} to deal with larger molecules we also report experiments on GEOM-DRUGS, the largest and most pharmaceutically relevant part of the GEOM dataset \citep{geom} — consisting of 304k drug-like molecules (average 44 atoms).  We report our results in Tab. \ref{tab:drugs} and compare with GeoDiff \citep{geodiff}, GeoMol \citep{geomol} and Torsional Diff. \citep{torsionaldiff}. Note again that all baseline approaches make strong assumptions about the geometric structure of molecules and model domain-specific characteristics like torsional angles of bonds. {\model} simply models the distribution of 3D coordinates of atoms without making any assumptions about the underlying structure. 

Results on Tab. \ref{tab:drugs} are where we see {\model} outperforms strong baseline approaches by substantial margins. All {\model} models achieve better performance than previous state-of-the-art Torsional Diff. model. On both recall and precision, {\model} of small size ({\model}-S) surpasses Torsional Diff. by approximately 15\%. This indicates that our proposed {\model} not only generates high-quality conformers that are close with ground truth but also covers a wide variety of conformers in the distribution. In addition, it is important to note that {\model} does not make any assumptions about the intrinsic geometric factors in conformers like torsional angles and thus provides a simple recipe to scale up the model. With growing number of parameters, larger {\model} constantly achieves better performance than smaller counterpart in all metrics. In particular, when compared with {\model}-S, {\model}-B shows approximately 15\% improvement on precision and even larger {\model}-L improves it by approximately 20\%. The experimental results demonstrate the power of scaling up proposed {\model} in better solving conformer generation problem. Since our proposed method simply operates on 3D atomic positions, it provides a straightforward recipe for scaling up the model. This sheds light on how scaling law could potentially benefit applications of deep generative models to scientific domains. 

In Fig. \ref{fig:cov_func_thres}, we further show a breakdown of the performance on GEOM-DRUGS of {\model} with different sizes vs. Torsional diffusion \citep{torsionaldiff} as a function of the threshold distance, as well as a function of the number of atoms in molecules. {\model} outperforms Torsional Diff. across the full spectrum of thresholds in both recall and precision. When looking at the break-down AMR on different number of atoms in Fig.~\ref{fig:cov_func_thres}(c), {\model} also demonstrates its superior performance for molecules of different sizes. It is indicated that {\model} better captures the fine intrinsic geometric structure of conformers and scaling up the model helps improve the performance of proposed model. Also, as the number of parameters increases, {\model} demonstrates better performance across all threshold levels in terms of both recall and precision. This provides further evidence on the performance gain from increasing the models size of {\model} which is designed to be scalable in a straightforward way.  We further investigate the ensemble properties of generated conformers in Appendix~\ref{app:ensemble}. Fig.~\ref{fig:drugs_examples} in the Appendix shows examples of {\model} generated conformers in GEOM-DRUGS.

\subsection{Generalization to GEOM-XL}

We now turn to the task of evaluating how well a model trained on GEOM-DRUGS transfers to unseen molecules with large numbers of atoms. Following \citet{torsionaldiff}, we use the GEOM-XL dataset, a subset of GEOM-MoleculeNet that contains 102 molecules with more than 100 atoms. Note that this evaluation not only tests the capacity of models to generalize to larger and more complex molecules but also serves as an out-of-distribution generalization experiment.

\begin{table}[t]
\setlength{\tabcolsep}{3pt}
    \centering
    \footnotesize
\begin{tabular}{l cc cc c}
& \multicolumn{2}{c}{AMR-P $\downarrow$} & \multicolumn{2}{c}{AMR-R $\downarrow$} & \# mols \\  
\toprule
& mean & median & mean & median &   \\
\midrule
GeoDiff & 2.92 & 2.62 & 3.35 & 3.15 & - \\
GeoMol & 2.47 & 2.39 & 3.30 & 3.14 & - \\
Tor. Diff. & 2.05 & 1.86 & \textbf{2.94} & 2.78 & - \\
{\model}-S & 2.22 & 1.97 & 3.17 & 2.81 & 102 \\
{\model}-B & 2.01 & 1.70 & 3.03 & 2.64 & 102 \\
{\model}-L & \textbf{1.97} & \textbf{1.60} & \textbf{2.94} & \textbf{2.43} & 102 \\
\midrule
Tor. Diff. (our eval) & 1.93 & 1.86 & 2.84 & 2.71 &	77 \\
{\model}-S & 2.02 & 1.87 & 2.9 & 2.69 & 77 \\
{\model}-B & 1.71 & 1.61 & 2.69 & 2.44 & 77 \\
{\model}-L & \textbf{1.64} & \textbf{1.51} & \textbf{2.57} & \textbf{2.26} & 77 \\
\bottomrule
\end{tabular}
\caption{Generalization results on GEOM-XL. 
}
\label{tab:xl}
\end{table}

In Tab. \ref{tab:xl} we report AMR for both precision and recall and compare with GeoDiff \citep{geodiff}, GeoMol \citep{geomol} and Torsional Diff. \citep{torsionaldiff}. In particular, when taking the numbers directly from \citet{torsionaldiff}, {\model}-B achieves better or comparable performance than Torsional Diff. Further, in running the checkpoint provided by Torsional Diff. and following their validation process we found that 25 molecules failed to be generated, this is due to the fact that Torsional Diff. generates torsional angles conditioned on the molecular graph $\mathcal{G}$ and the local structures obtained from RDKit. And RDKit can fail to find local structures and Torsional Diff. cannot generate conformers in these cases. In our experiments with the same 77 molecules in GEOM-XL from our replica, {\model} surpasses Torsional Diff. by a large margin. Leveraging little or no inductive bias in modeling molecular conformers, our proposed method is adaptable to wider variety of data. 

Results also show that scaling up the model size to {\model}-L further improves the generalizability to large and unseen molecules in GEOM-XL. In our replica, {\model}-L demonstrates better performance than smaller model counterparts (\ie {\model}-S and {\model}-B) and surpasses Torsional Diff. by a large margin. The results highlight the generalizability of {\model} to large and complex molecules, which may shed a light on pre-training molecular conformer generation model.

\subsection{Why does {\model} generalize?}

\begin{figure}[t]
\centering
\setlength{\tabcolsep}{1pt}
    \begin{tabular}{cc}
    \centering
    \includegraphics[width=0.24\textwidth]{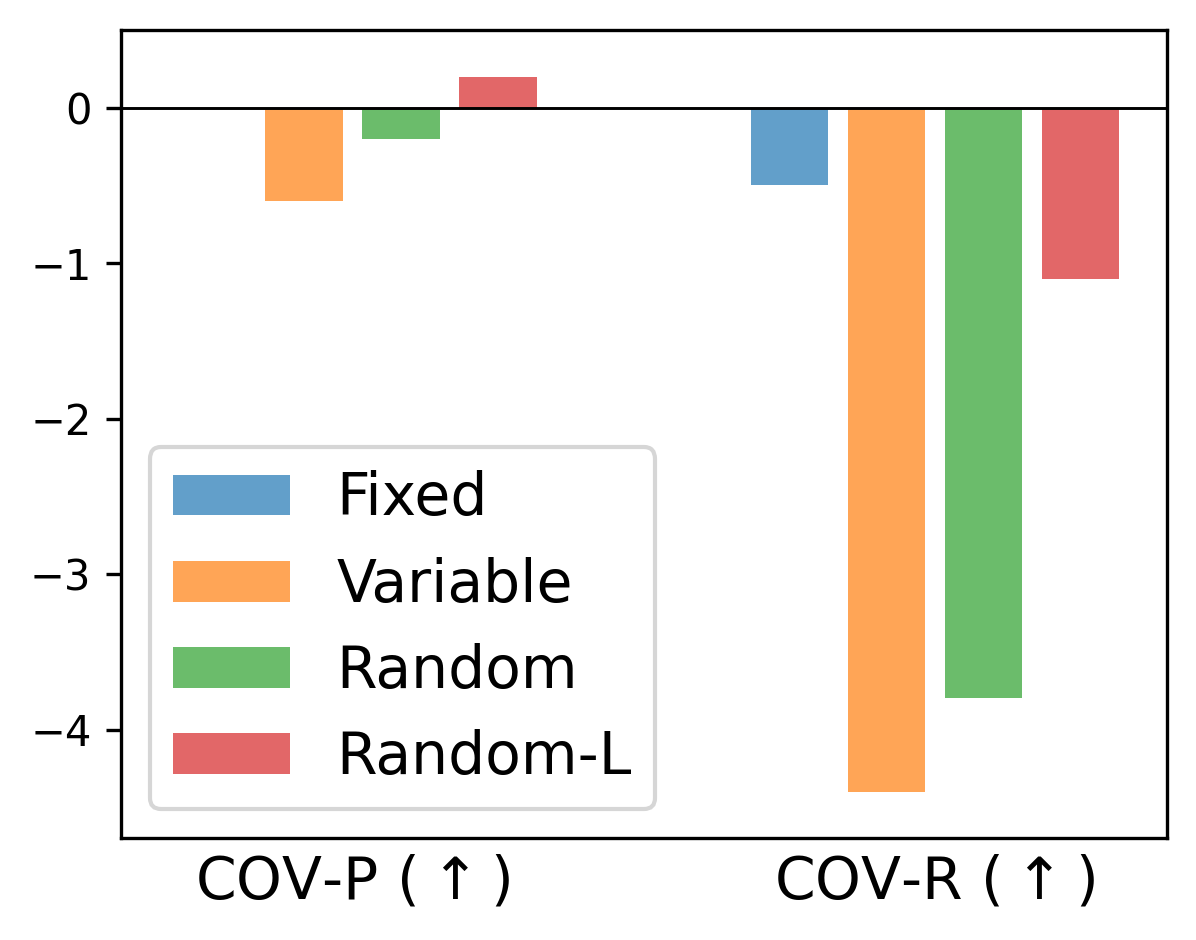} & \includegraphics[width=0.24\textwidth]{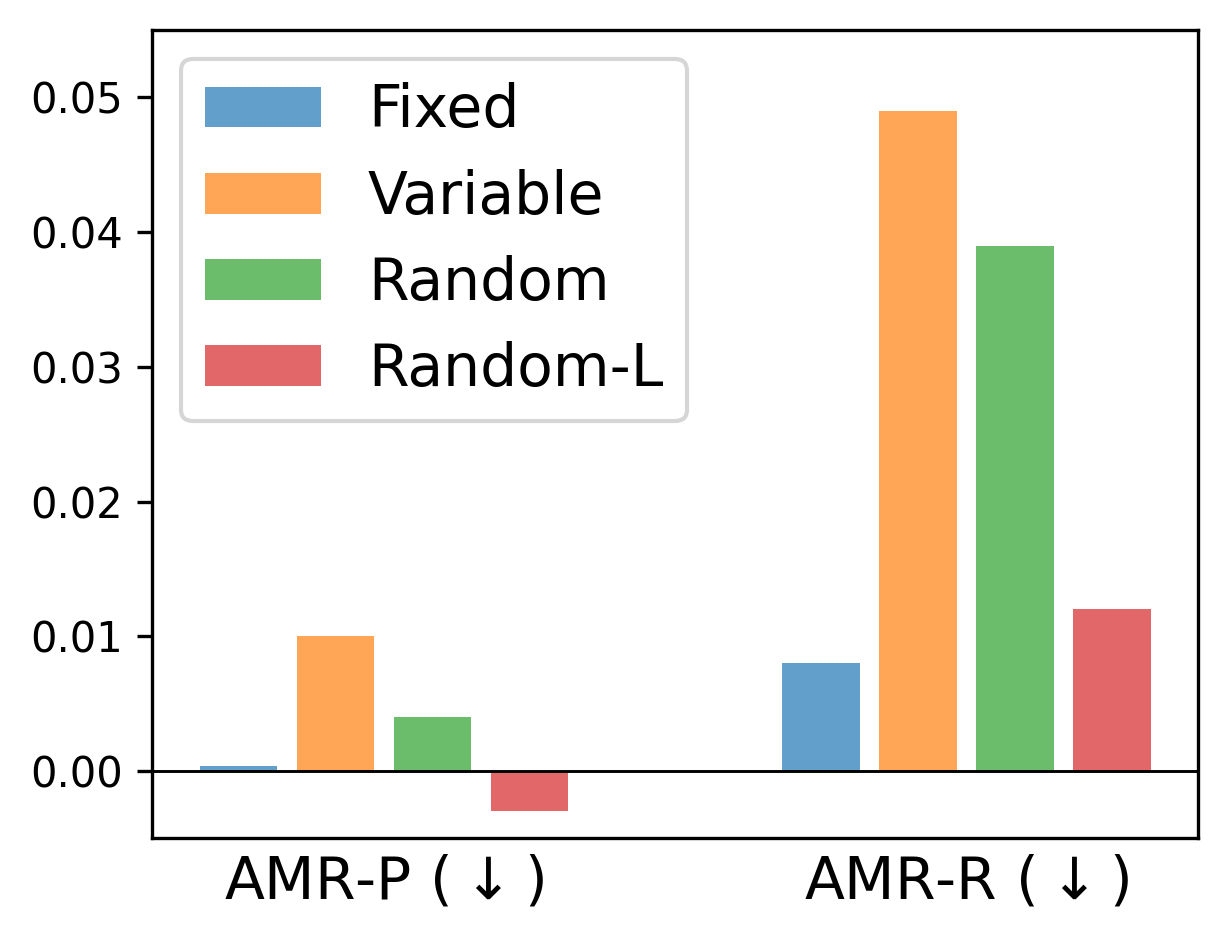} \\
        (a) & (b)
    \end{tabular}
    \caption{(a) Mean Coverage and (b) mean AMR of different rotation augmentation strategies on GEOM-QM9 when compared with training on original dataset. }
    \label{fig:rot}
\end{figure}

A natural question to ask is why does {\model} generalize given its non-equivariant design. To answer this question we devised an experiment to understand if conformers in training and validation sets share a canonical coordinate system. In our experiment we apply different rotation transformations to GEOM-QM9 and train {\model} on this transformed training set, while keeping the validation set unchanged. Three rotation transformations are investigated: 1) ``Fixed" applies a single random rotation to all conformers, 2) ``Variable" applies a different rotation to each conformer and keeps it through training, 3) ``Random" applies a different random rotation to each conformer in each training epoch. Fig.~\ref{fig:rot} shows the results in these different settings. 

Not surprisingly, applying a fixed rotation to the training set minimally affects performance. This is because a fixed rotation does not break relative SO(3) relations between conformers in the training set. However, rotating each conformer independently once during training (\eg ``Variable") negatively impacts performance. This finding points to the fact that the DFT simulations used to generate the data might be implicitly encoding a canonical coordinate system, which affects generalization if broken. Finally, applying a random rotation to each conformer on each training epoch forces the model to be invariant to any coordinate system, which is a more challenging task. Notably, though randomly rotating each conformer leads to worse results in recall, the performance drop is still marginal. Finally, by  training a bigger model on this randomly rotated training set (\ie Random-L) we can recover most of the performance gap in comparison with training on the original dataset.


These experiments show that inductive biases like roto-equivariance can be traded for scale in general-purpose models. Our results highlight that a domain-agnostic model at scale can achieve better performance than intricate models with strong domain-specific inductive biases in molecular conformer generation. We hope our findings will inspire the community to develop simple models that are prone to benefit from ``scaling laws" especially when taking into account the fast growth of available scientific data. 


\subsection{Sampling}
\label{sect:ddim}

\begin{figure}[t]
\centering
\setlength{\tabcolsep}{1pt}
    \begin{tabular}{cc}
    \centering
    \includegraphics[width=0.24\textwidth]{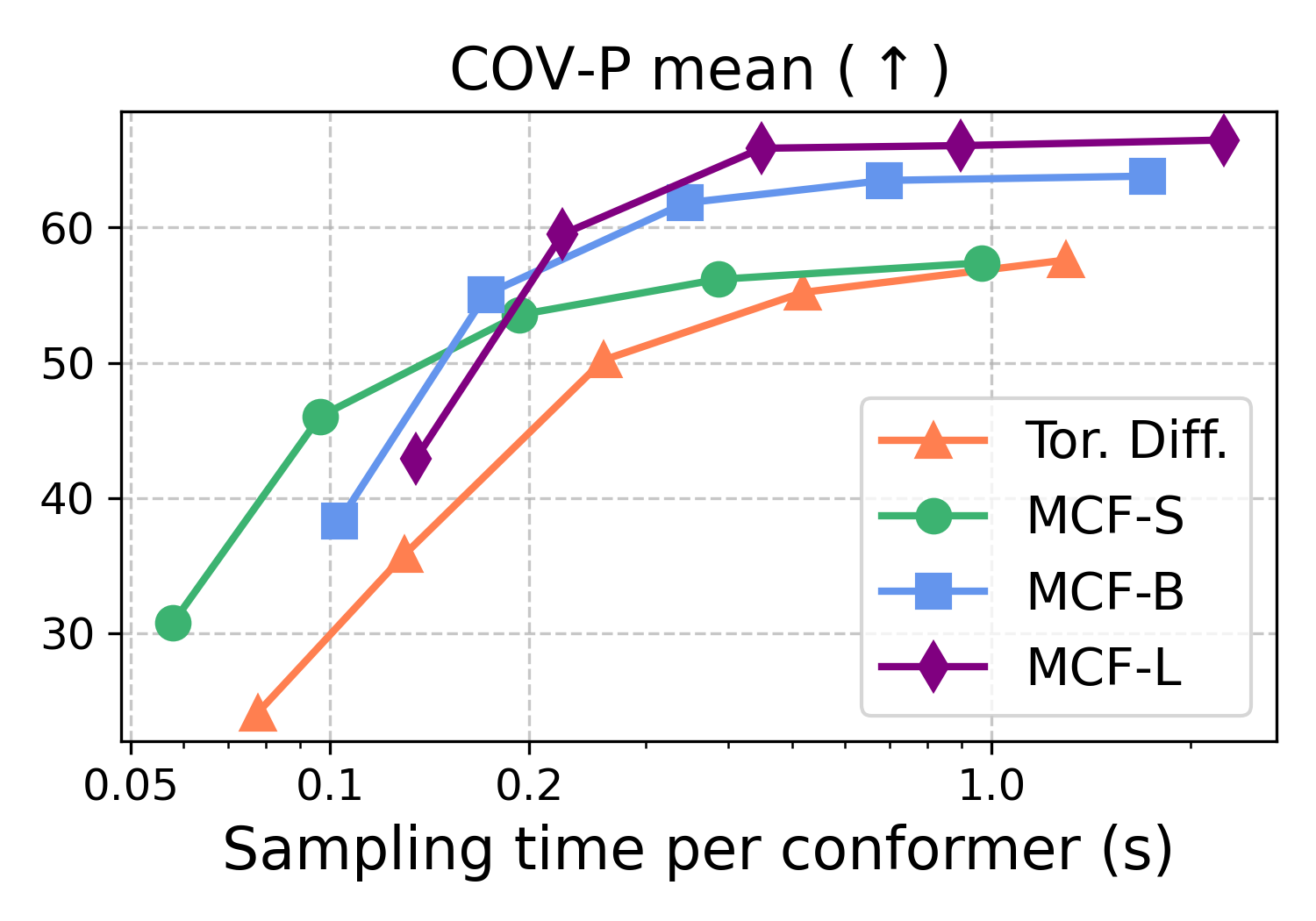} & \includegraphics[width=0.24\textwidth]{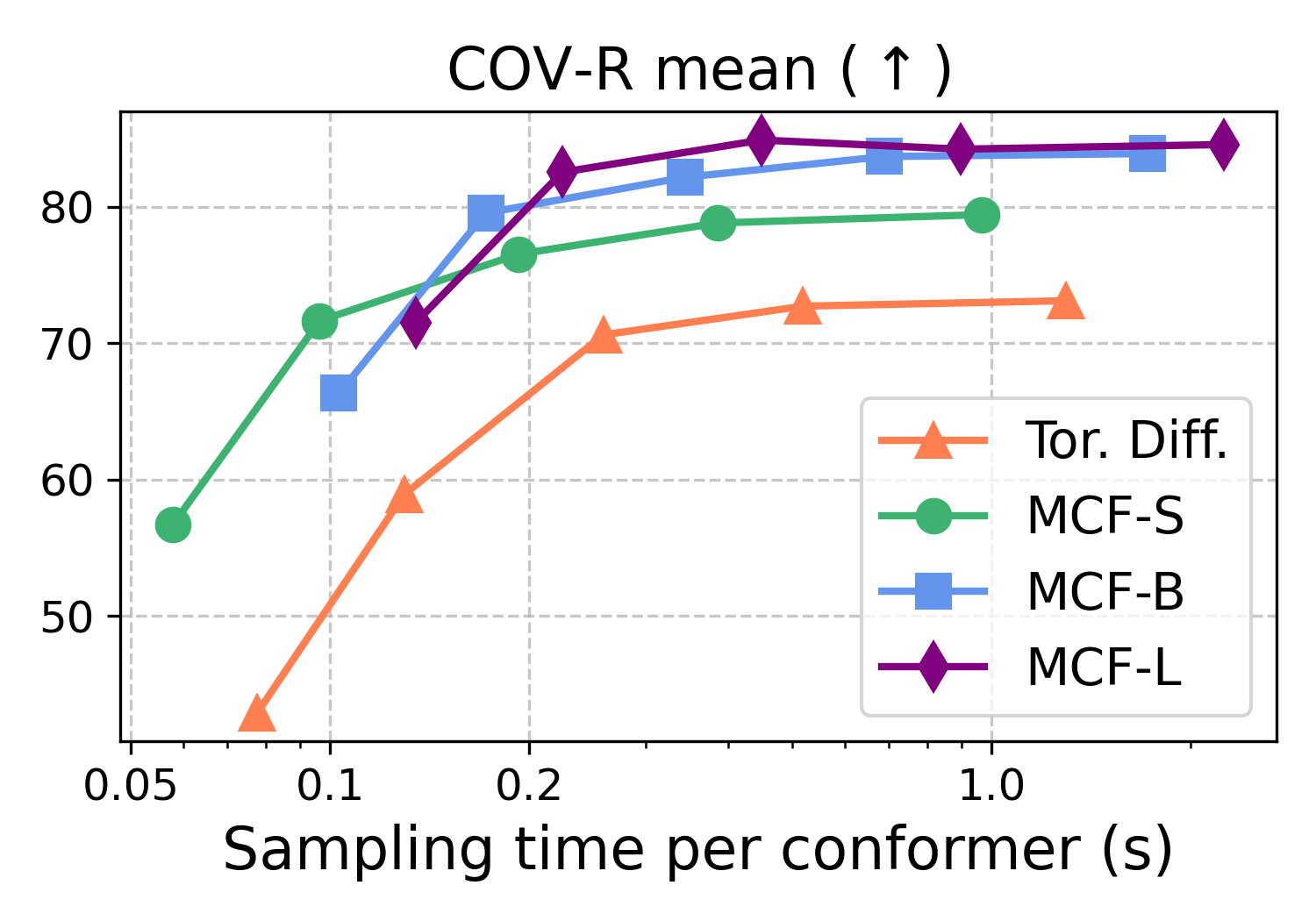} \\
    (a) & (b) \\
    \includegraphics[width=0.24\textwidth]{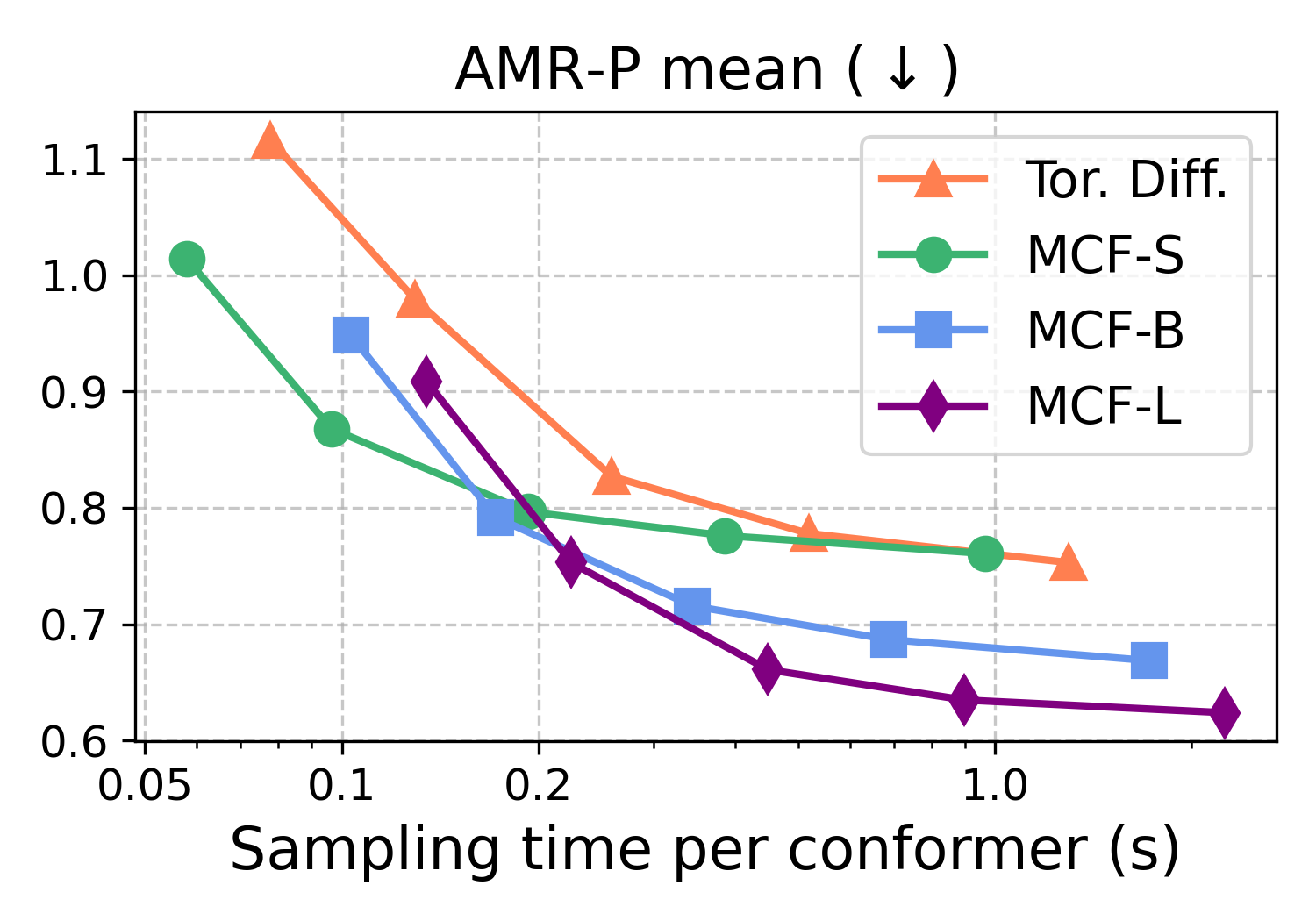} &
    \includegraphics[width=0.24\textwidth]{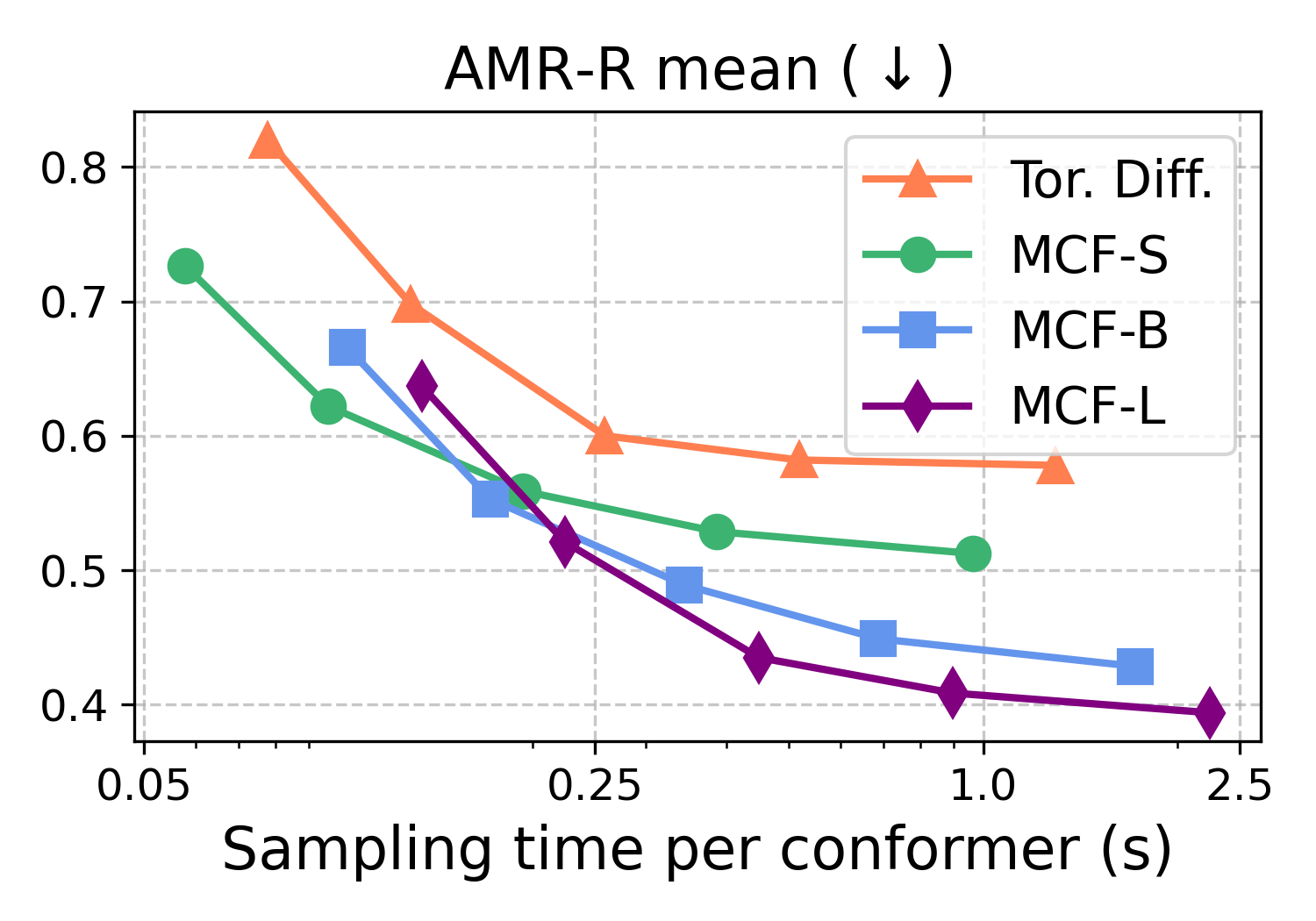} \\
    (c) & (d)
    \end{tabular}
    \caption{Inference wall clock time v.s. (a) precision coverage, (b) recall coverage, (c) precision AMR, and (d) recall AMR with Torsional Diff. and our {\model}. }
    \label{fig:ddim}
\end{figure}

In this section, we investigate the performance of our {\model} under limited computation budget in inference. To this end, we report COV and AMR of {\model} with respect to different wall-clock sampling times. DDIM \citep{ddim}, an efficient sampler, is applied, which uses a significantly smaller number of sampling steps than vanilla DDPM (\ie 1000 sample steps). Specifically, we sample conformers with 3, 5, 10, 20, and 50 sampling steps with DDIM and compare the performance as well as inference time with Torsional Diff. \citep{torsionaldiff}. All models are benchmarked on a single A100 GPU for comparison. It is shown that {\model} is more efficient than Torsional Diffusion, which means that for the same inference time in seconds, {\model} always outperforms Torsional Diffusion across all metrics and all model sizes. Notably, due to application of equivariant operations, Torsional Diff. can be time consuming in inference. Even when using a very limited sampling steps (\eg 5 steps) {\model} achieves comparable or better COV and AMR than Torsional Diff. using 50 sampling steps, while being more efficient in wall-clock inference time. {\model} with different sizes achieves Pareto frontier of performance (\ie COV and AMR) and sampling efficiency compared with Torsional Diff. This further indicates rotational or translational equivariance may not be a strong requirement while simple and scalable framework like {\model} can own the merits in efficiency. Examples of {\model} sampled conformers with different sampling steps can be found in Fig.~\ref{fig:ddim_samples} in the Appendix.

\section{Conclusions}

In this paper we introduced {\model}, where we formulate the problem of molecular conformer generation as learning a diffusion model over functions on molecular graphs. {\model} achieves state-of-the-art performance across different  molecular generation benchmarks, surpassing models with hard-coded inductive biases by a large margin. Notably, {\model} uses general-purpose Transformer-based score network rather than a model designed with specific inductive biases for molecules. {\model} achieves superior results without explicitly modeling geometric properties of molecules like torsional angles, which makes it simpler to understand and scale. We believe {\model} represents an exciting first step for future research on scaling conformer generation to proteins and other macro molecular structures. We hope our work serves as a reminder to the community to carefully consider the interplay between baking inductive biases in architectures while also considering the benefits of efficient and scalable approaches.

\section*{Impact Statement}

This paper introduces a novel diffusion model based approach to generate molecular conformers, contributing to the advancement of computational chemistry and molecular modeling. The impact of this work extends to various scientific domains, enabling more accurate predictions of molecular structures and behaviors. The potential societal implications lie in the acceleration of drug discovery, materials science, and environmental studies, offering efficient solutions to complex molecular design challenges. 

\bibliography{main}
\bibliographystyle{icml2024}

\newpage
\appendix
\onecolumn

\section{Appendix}

\subsection{Limitations and Future Work}
\label{app:limitations}

While {\model} shows competitive performance in molecular conformer generation, it does encounter limitations and potential improvements for future explorations. One limitation is that our proposed method is computationally expensive. Extensive computations first stem from the Transformer-based \citep{transformers} score network. In {\model}, we use a PerceiverIO \citep{perceiverio} as score network, an efficient Transformer that allows for sub-quadratic compute, as well as FlashAttention \citep{flashattention} in implementation. Other efficient Transformer architectures and tricks like \citet{jabri2022scalable} can be used to improve training efficiency. The other factor is computational cost during inference. In {\model}, we iterate 1000 timesteps to sample a conformer following DDPM \citep{ddpm}. Experiments in Section~\ref{sect:ddim} show that efficient sampling strategies, \ie DDIM \citep{ddim}, can help significantly increase inference efficiency while maintain high-quality in sampled conformers. Other efficient variants of diffusion models like consistency model \citep{consistency_models} as well as distillation approaches \citep{tract} may be adapted to further decrease the sampling to single step. Also, recent works have demonstrated that diffusion generative model can generate samples following Boltzmann distributions when provided with Boltzmann-distributed training data \cite{arts2023two}. Driven by this, our proposed {\model} can be adapted to generate molecular conformers that follow Boltzmann distributions when trained with corresponding data. Besides, recent flow matching generative model \citep{lipman2022flow} provides the flexibility of mapping between arbitrary distributions and access to exact log-likelihood estimation. Integrating flow matching framework could help sample molecular conformer from Boltzmann distribution instead of standard Gaussian. Some recent works \citep{flam2023language, o20243d, gruver2022lie} also show that expressive models can learn equivariance from data, but they have not thoroughly investigated molecular conformer generation.

Another limitation could be how well {\model} performs in low data regime. The propsed method may not perform as well as conformer generation when applied to problems with limited data or related to sequential problems like molecular dynamics (MD) simulations. In future work, we plan to extend {\model} to conditional inference. For example, molecular docking can be formulated as conformer generation problem conditioned on proteins \citep{diffdock}. Also, current framework can be expanded to \emph{de novo} drug designs where no molecule information is provided \citep{edm}. Besides, scaling up our model to large molecules, like proteins, can be of great interest. {\model} by nature provides the flexibility to generate from partially observed sample, which can be suitable for designing proteins with known functional motifs \citep{rfdiffusion}.

\subsection{Implementation details}
\label{app:implementation_details}
In this section we describe implementation details for all our experiments. We also provide hyper-parameters and settings for the implementation of the score field network $\epsilon_\theta$ and compute used for each experiment in the paper. In our experiments, we split GEOM-QM9 and GEOM-DRUGS randomly based on molecules into train/validation/test (80\%/10\%/10\%). At the end, for each dataset, we report the performance on 1000 test molecules. Thus, the splits contain 106586/13323/1000 and 243473/30433/1000 molecules for GEOM-QM9 and GEOM-DRUGS, respectively. We follow the exact same training splits for all baselines \cite{geomol, torsionaldiff}.

\subsubsection{Score Field Network implementation details}

The time-step $t$ is incorporated into the score computation by concatenating a positional embedding representation of $t$ to the context and query sets. The specific PerceiverIO settings used in all quantitatively evaluated experiments are presented in Tab.~\ref{tab:perceiver_io_hparams}. An AdamW \citep{adamw} optimizer is employed during training with a learning rate of $1e-4$. Cosine learning rate decay is deployed with 30K warmup steps. We use EMA with a decay of $0.999$. Models are trained for 300K steps on GEOM-QM9 and 750K steps on GEOM-DRUSG. All models use an effective batch size of 512. A modified version of the publicly available repository is used for PerceiverIO \footnote{\url{https://huggingface.co/docs/transformers/model_doc/perceiver}}. Since molecules have different number of atoms, we set the number of context and query sets as the number of atoms during training and inference.



\begin{table*}[h]
\scriptsize
    \centering
    \footnotesize
    \setlength{\tabcolsep}{10pt}
    \begin{tabular}{l c c c}
    \toprule
        Hyper-parameter & Small & Base & Large \\
        \midrule
        \texttt{num\_freq\_pos\_embed} & $128$ & $128$ & $128$ \\
        \texttt{num\_latent}           & $128$ & $512$ & $1024$  \\ 
        \texttt{d\_latennt}            & $256$ & $512$ & $1024$ \\
        \texttt{d\_model}              & $512$ & $1024$ & $1024$  \\
        \texttt{num\_enc\_block}       & $6$ & $8$ & $12$ \\
        \texttt{num\_dec\_block}       & $2$ & $2$ & $2$ \\
        \texttt{num\_self\_attn\_per\_block} & $2$ & $2$  & $2$ \\
        \texttt{num\_self\_attn\_head}   & $4$ & $4$ & $8$ \\
        \texttt{num\_cross\_attn\_head}  & $4$ & $4$ & $8$ \\
        \midrule
        \texttt{\# param} & 13M & 64M & 242M \\
        \bottomrule
    \end{tabular}
    \caption{Hyperparameters and settings for {\model} on different datasets.}
    \label{tab:perceiver_io_hparams}
\end{table*}

\subsubsection{Atomic Features}
\label{app:atomic}

We include atomic features alongside the graph Laplacians to model the key descriptions of molecules following previous works \cite{geomol, torsionaldiff}. Detailed features are listed in Tab.~\ref{tab:atomic}. The atomic features are concatenated with graph Laplacian eigenvectors in both context and query inputs. 

\begin{table*}[t!]
  \centering
  \small
  \begin{tabular}{l l l}
    \toprule
    Name & Description & Range \\
    \midrule
    \texttt{atomic} & Atom type & one-hot of 35 elements in dataset \\
    \texttt{degree} & Number of bonded neighbors & $\{x:0 \leq x \leq 6, x \in \mathbbm{Z}\}$ \\
    \texttt{charge} & Formal charge of atom & $\{x:-1 \leq x \leq 1, x \in \mathbbm{Z}\}$ \\
    \texttt{valence} & Implicit valence of atom & $\{x:0 \leq x \leq 6, x \in \mathbbm{Z}\}$ \\
    \texttt{hybrization} & Hybrization type & \{sp, sp\textsuperscript{2}, sp\textsuperscript{3}, sp\textsuperscript{3}d, sp\textsuperscript{3}d\textsuperscript{2}, other\} \\
    \texttt{aromatic} & Whether on a aromatic ring & \{True, False\} \\
    \texttt{num\_rings} & number of rings atom is in & $\{x:0 \leq x \leq 3, x \in \mathbbm{Z}\}$ \\
    \bottomrule
  \end{tabular}
  \caption{Atomic features included in {\model}.}
  \label{tab:atomic}
\end{table*}

\subsubsection{Compute}
For GEOM-QM9, we train models using a machine with 4 Nvidia A100 GPUs using precision BF16. For GEOM-DRUGS, we train models using precision FP32, where {\model}-B is trained with 8 Nvidia A100 GPUs and {\model}-L is trained with 16 Nvidia A100 GPUs. 

\subsubsection{Evaluation Metrics}
\label{app:metrics}

Following previous works \cite{geodiff, geomol, torsionaldiff}, we apply Average Minimum RMSD (AMR) and Coverage (COV) to measure the performance of molecular conformer generation. Let $C_g$ denote the sets of generated conformations and $C_r$ denote the one with reference conformations. For AMR and COV, we report both the Recall (R) and Precision (P). Recall evaluates how well the model locates ground-truth conformers within the generated samples, while precision reflects how many generated conformers are of good quality. The expressions of the metrics are given in the following equations:
\begin{equation}
    \text{AMR-R}(C_g, C_r) = \frac{1}{|C_r|} \sum_{\mathbf{R} \in C_r} \min_{\mathbf{\hat{R}} \in C_g} \text{RMSD}(\mathbf{R}, \mathbf{\hat{R}}),
\label{eq:amrr}
\end{equation}
\begin{equation}
    \text{COV-R}(C_g, C_r) = \frac{1}{|C_r|} |\{ \mathbf{R} \in C_r | \text{RMSD}(\mathbf{R}, \mathbf{\hat{R}}) < \delta, \mathbf{\hat{R}} \in C_g \}|,
\label{eq:covr}
\end{equation}
\begin{equation}
    \text{AMR-P}(C_r, C_g) = \frac{1}{|C_g|} \sum_{\mathbf{\hat{R}} \in C_g} \min_{\mathbf{R} \in C_r} \text{RMSD}(\mathbf{\hat{R}}, \mathbf{R}),
\label{eq:amrp}
\end{equation}
\begin{equation}
    \text{COV-P}(C_r, C_g) = \frac{1}{|C_g|} |\{ \mathbf{\hat{R}} \in C_g | \text{RMSD}(\mathbf{\hat{R}}, \mathbf{R}) < \delta, \mathbf{R} \in C_r \}|,
\label{eq:covp}
\end{equation}
where $\delta$ is a threshold. In general, a lower AMR scores indicate better accuracy and a higher COV score indicates a better diversity for the generative model. Following \cite{torsionaldiff}, $\delta$ is set as $0.5 \angstrom$ for GEOM-QM9 and $0.75 \angstrom$ for GEOM-DRUGS.

\subsection{Additional experiments}
\label{app:additional_experiments}

In this section we include additional experiments ablating architecture choices, as well as prediction the ensemble properties of generated conformers.

\subsubsection{Ablation experiments}

In this section we provide an ablation study over the key design choices of {\model}. We run all our ablation experiments on the GEOM-QM9 dataset following the settings in GeoMol \citep{geomol} and Torsional Diff. \citep{torsionaldiff}. In particular we study: (i) how does performance behave as a function of the number of Laplacian eigenvectors used in $\varphi(v)$. (ii) How does the model perform without atom features (\eg how predictable conformers are given only the graph topology, without using atom features). Results in Tab. \ref{tab:ablation} show that the graph topology $\mathcal{G}$ encodes a surprising amount of information for sampling reasonable conformers in GEOM-QM9, as shown in row 2. In addition, we show how performance of {\model} changes as a function of the number of eigen-functions $k$. Interestingly, with as few as $k=2$ eigen-functions {\model} is able to generate consistent accurate conformer.


\subsubsection{Architectural choices}

To further investigate the design choices of architecture in proposed {\model}, we include additional experiments on GEOM-QM9 as shown in Tab.~\ref{tab:ablation}. To investigate the effectiveness of using Laplacian eigenvectors from LBO eigen-decomposition as positional embedding, we leverage SignNet \citep{signnet} as the positional embedding, which explicitly models symmetries in eigenvectors. Using SignNet does not benefit the performance when compared with the standard {\model}. Though adding edge attributes in SignNet achieves better performance than SignNet alone, the performance is still not rival. Also, it's worth mentioning that SignNet includes graph neural networks \citep{gin} and Set Transformer \citep{set_transformer} which makes training less efficient. 

In addition, we also report results using a vanilla Transformer encoder-decoder (TF) \citep{transformers} as the backbone instead of PerceiverIO (PIO) \citep{perceiverio}. TF-base model contains 6 encoder layers and 6 decoder layers with 4 attention heads while TF-large contains 12 encoder layers and 12 decoder layers. The number of parameters match approximately with base and large sized PerceiverIO investigated in this work. Tab.~\ref{tab:ablation} shows that TF-base is performing significantly worse than PIO-base with similar number of parameters. When increasing the model size, TF-large achieves on par performance as PIO-base, which validates the design choice of architecture in {\model}.

\begin{table*}[!h ]
\setlength{\tabcolsep}{3pt}
    \centering
    \small
\begin{tabular}{llll cc cc cc cc}
& & & & \multicolumn{4}{c}{Precision} & \multicolumn{4}{c}{Recall} \\   
\toprule
& & & & \multicolumn{2}{c}{COV $\uparrow$} & \multicolumn{2}{c}{AMR $\downarrow$}  & \multicolumn{2}{c}{COV $\uparrow$} & \multicolumn{2}{c}{AMR $\downarrow$} \\
\toprule
$k$ & atom feat. & PE & backbone & mean & median & mean & median & mean & median & mean & median \\
\midrule
28 & YES & LBO & PIO-base & 95.00 & 100.00 & 0.103 & 0.044 & 93.67 & 100.00 & 0.119 & 0.055 \\
28 & NO  & LBO & PIO-base & 90.70 & 100.00 & 0.187 & 0.124 & 79.82 & 93.86 & 0.295 & 0.213 \\
16 & YES & LBO & PIO-base & 94.87 & 100.00 & 0.139 & 0.093 & 87.54 & 100.00 & 0.220 & 0.151 \\
8  & YES & LBO & PIO-base & 94.28 & 100.00 & 0.162 & 0.109 & 84.27 & 100.00 & 0.261 & 0.208 \\
4  & YES & LBO & PIO-base & 94.57 & 100.00 & 0.145 & 0.093 & 86.83 & 100.00 & 0.225 & 0.151 \\
2  & YES & LBO & PIO-base & 93.15 & 100.00 & 0.152 & 0.088 & 86.97 & 100.00 & 0.211 & 0.138 \\
28 & YES & SignNet & PIO-base & 94.10 & 100.00 & 0.153 & 0.098 & 87.50 & 100.0 & 0.222 & 0.152 \\
28 & YES & SignNet\textsubscript{attr} & PIO-base & 95.30 & 100.00 & 0.143 & 0.091 & 90.20 & 100.00 & 0.197 & 0.135 \\
28 & YES & LBO & TF-base & 94.92 & 100.00 & 0.131 & 0.083 & 89.33 & 100.00 & 0.194 & 0.132 \\
28 & YES & LBO & TF-large & 95.49 & 100.00 & 0.110 & 0.061 & 93.48 & 100.00 & 0.135 & 0.073 \\
\bottomrule
\end{tabular}
\caption{Ablation study with different network architectures on GEOM-QM9.}
\label{tab:ablation}
\end{table*}

\subsubsection{Ensemble properties}
\label{app:ensemble}

To fully assess the quality of generated conformers we also compute chemical property resemblance between the synthesized and the authentic ground truth ensembles. We select a random group of 100 molecules from the GEOM-DRUGS and produce a minimum of $2K$ and a maximum of $32$ conformers for each molecule following \citep{torsionaldiff}. Subsequently, we undertake a comparison of the Boltzmann-weighted attributes of the created and the true ensembles. To elaborate, we calculate the following characteristics using xTB (as documented by \citep{xtb}): energy ($E$), dipole moment ($\mu$), the gap between HOMO and LUMO ($\Delta \epsilon$), and the lowest possible energy, denoted as $E_{\text{min}}$. Since we don't have the access to the exact subset of DRUGS used in \cite{torsionaldiff}, we randomly pick three subsets and report the averaged and standard deviation over three individual runs with different random seeds. The results are listed in Tab.~\ref{tab:ensemble}. Our model achieves the lowest error on $E_{\text{min}}$ when compared with other baselines, which demonstrates that {\model} succeeds at generating stable conformers that are very close to the ground states. This could root from the fact that {\model} doesn't rely on rule-based cheminfomatics methods and the model learns to better model stable conformers from data. Besides, {\model} achieves competitive performance on $\mu$ and $\Delta \epsilon$. However, the error of $E$ is high compared to the rest of approaches, meaning that though {\model} performs well in generating samples close to ground states, it may also generate conformers with high energy that are not plausible in the dataset. 

\begin{table}[t]
\setlength{\tabcolsep}{3pt}
    \centering
    \small
\begin{tabular}{l cc cc}
\toprule
& $E$ & $\mu$ & $\Delta \epsilon$ & $E_{\text{min}}$ \\
\midrule
OMEGA & 0.68 & 0.66 & 0.68 & 0.69 \\
GeoDiff & 0.31 & 0.35 & 0.89 & 0.39 \\
GeoMol & 0.42 & 0.34 & 0.59 & 0.40 \\
Tor. Diff. & \textbf{0.22} & 0.35 & \textbf{0.54} & 0.13 \\
{\model} & 0.68$\pm$0.06 & \textbf{0.28}$\pm$\textbf{0.05} & 0.63$\pm$0.05 & \textbf{0.04}$\pm$\textbf{0.00}\\
\midrule
Tor. Diff. (our eval) & 3.07$\pm$2.32 & 0.61$\pm$0.38 & 1.71$\pm$1.69 & 4.11$\pm$7.91 \\
{\model} & 1.00$\pm$0.70 & 0.44$\pm$0.36 & 1.32$\pm$1.40 & 1.16$\pm$2.02 \\
\bottomrule
\end{tabular}
\caption{Median averaged errors of ensemble properties between sampled and generated conformers ($E$, $\Delta \epsilon$, $E_{\text{min}}$ in kcal/mol, and $\mu$ in debye). }
\label{tab:ensemble}
\end{table}


\begin{figure*}[t]
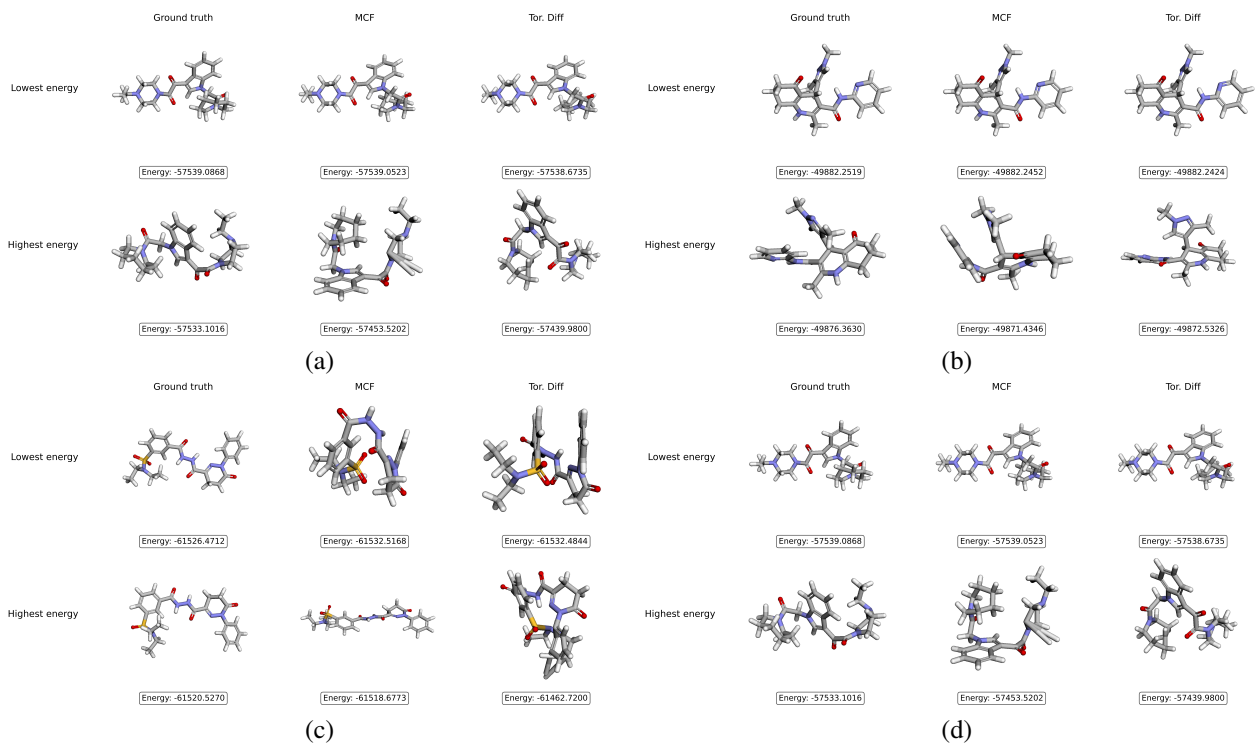

\setlength{\tabcolsep}{1pt}
    \begin{tabular}{ccc}
    \includegraphics[width=0.49\textwidth]{figure/ensemble_match1.jpg} & \includegraphics[width=0.49\textwidth]{figure/ensemble_match2.jpg} \\
    (a) & (b) \\
    \includegraphics[width=0.49\textwidth]{figure/ensemble_match3.jpg} & \includegraphics[width=0.49\textwidth]{figure/ensemble_match4.jpg} \\
    (c) & (d) \\
    \end{tabular}
    \caption{Examples of conformers with lowest and highest energies in ground truth, {\model} samples, and Torsional Diff. samples for different molecules.}
    \label{fig:ensemble_conf}
\end{figure*}


To further evaluate the performance on ensemble properties, we randomly pick 10 molecules from test set of GEOM-DRUGS and compare {\model} with our replica Torsional Diff. on the subset as shown in the last two rows of Tab.~\ref{tab:ensemble}. We use the checkpoints from the public GitHub repository\footnote{\url{https://github.com/gcorso/torsional-diffusion}} of Torsional Diff. to sample conformers. Unlike previous setting which only sample 32 conformers, we sample $2K$ conformers for a molecule with $K$ ground truth conformers. We report the average and standard deviation of errors over the 10 molecules. It is indicated that {\model} generates samples with ensemble properties that are closer to the ground truth. Fig.~\ref{fig:ensemble_conf} shows the conformers with lowest and highest energy in ground truth, {\model} samples, and Torsional Diff. samples. 

\subsection{Continuous conformers}
\label{sect:conf_field}

\begin{figure*}[!htb]
    \centering
    \includegraphics[width=\textwidth]{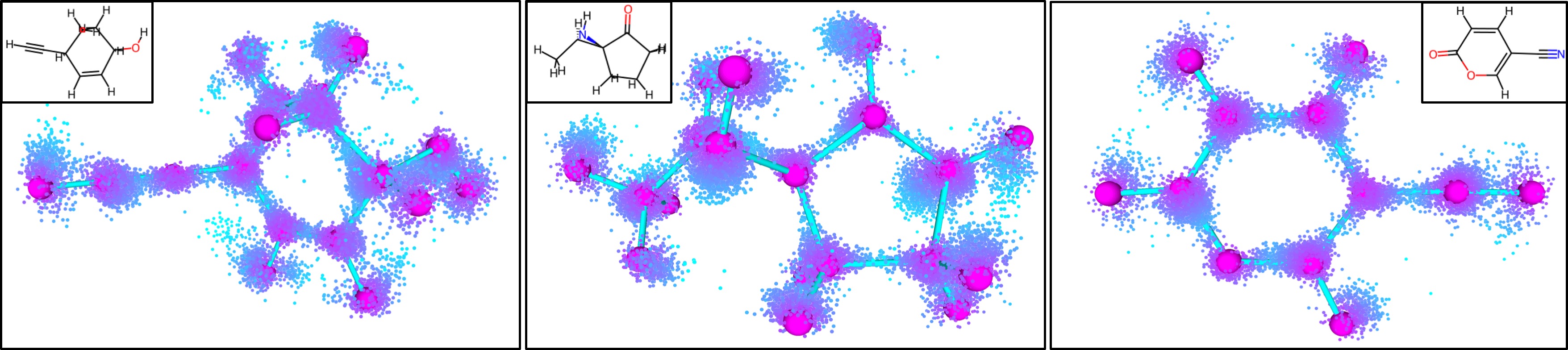}
    \caption{Continuously evaluating generated conformer fields for different molecules in GEOM-QM9.}
    \label{fig:conformer_field}
\end{figure*}

Molecular conformers are defined as discrete atomic positions in 3D Euclidean space. Since {\model} encodes continuous conformer field, it can be \textit{continuously evaluated} in $\mathcal{G}$, which maps arbitrary points in $\mathcal{G}$ to 3D positional space $\mathbbm{R}^3$. In order to do this, for a point $p$ in a bond connecting atoms $(v_i, v_j)$ we linearly interpolate the Laplacian eigenvector representation of it's endpoints $\varphi(p) = \alpha \varphi(v_i) + (1-\alpha)\varphi({v_j})$, we then feed this interpolated Laplacian eigenvector into the model to sample its 3D position in the conformer field.  We visualize results in Fig. \ref{fig:intro} and \ref{fig:conformer_field}. We generated this visualizations an {\model} model trained on GEOM-QM9 without atom features. Note that while {\model} is never trained on points along molecular bonds, it manages to generate plausible 3D positions for such points.

Here, the experiment conceptually investigates the flexibility of defining conformer generation problem as a field. It is shown that {\model} can generate feasible conformers even when the input interpolated eigenfunctions have never been seen during training. Such that {\model} is not over-fitted to certain eigenfunctions and learns to generate distributional aspects of atomic positions purely from correlations in training data. Also, when provided molecular conformer data with distribution of electron density from Quantum Monte Carlo methods \citep {qmc}, {\model} may be extended to predict electron density beyond atomic positions in future works as well. We recognize this is highly speculative and needs further empirical investigation to substantiate in future works. 

\subsection{Additional visualization}

\begin{figure*}[t]
\setlength{\tabcolsep}{1pt}
    \begin{tabular}{cc}
    \centering
    \includegraphics[width=0.98\textwidth]{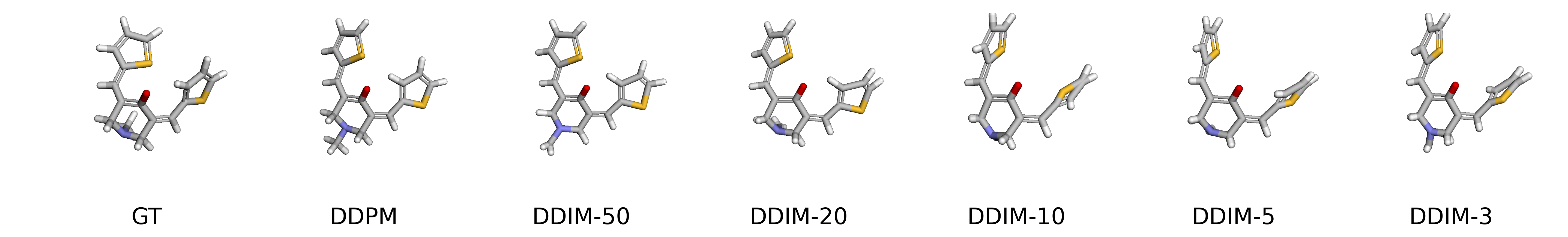} & \\
    \includegraphics[width=0.98\textwidth]{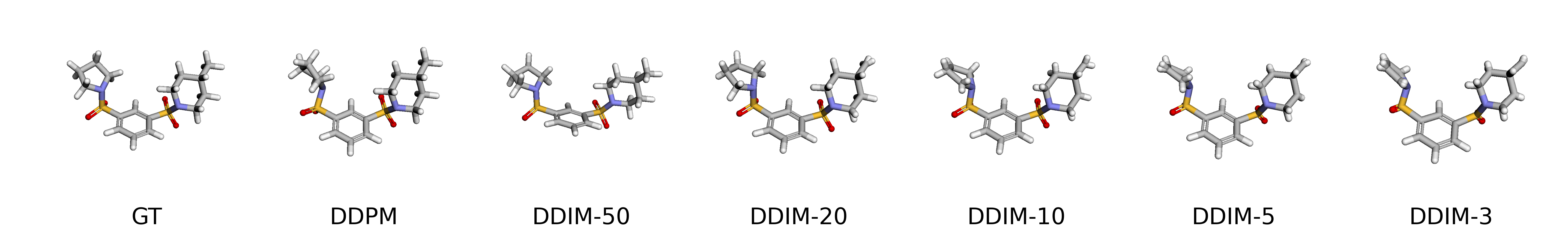} & \\
    \includegraphics[width=0.98\textwidth]{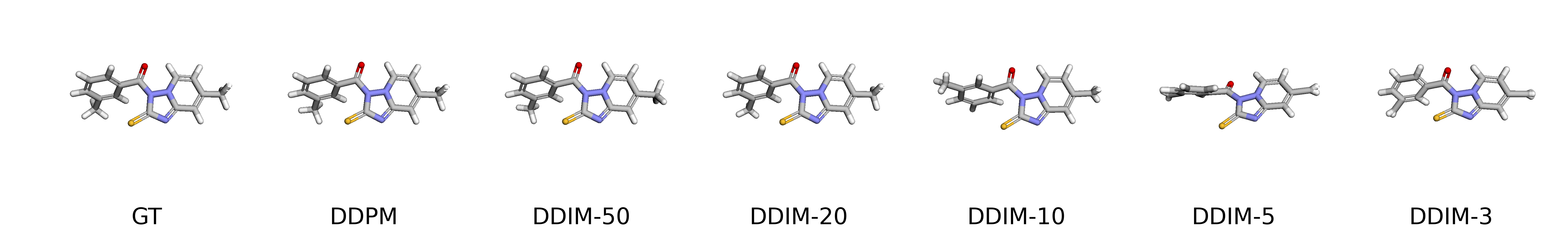} & \\
    \includegraphics[width=0.98\textwidth]{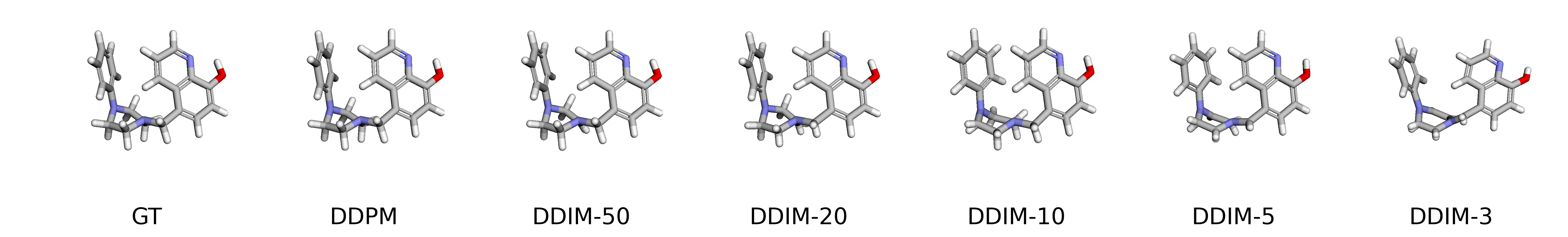} & \\
    \includegraphics[width=0.98\textwidth]{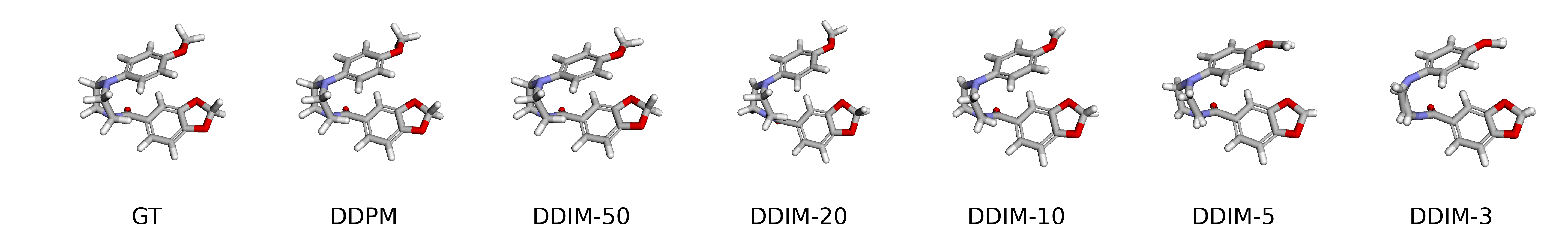} & \\
    \includegraphics[width=0.98\textwidth]{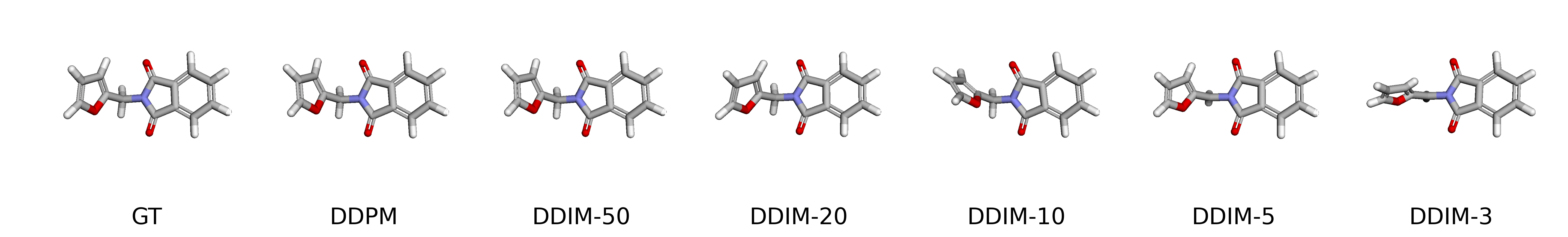} & \\
    \end{tabular}
    \caption{Examples of conformers with different sampling steps.}
    \label{fig:ddim_samples}
\end{figure*}

\begin{figure*}
    \centering
    \includegraphics[width=\textwidth]{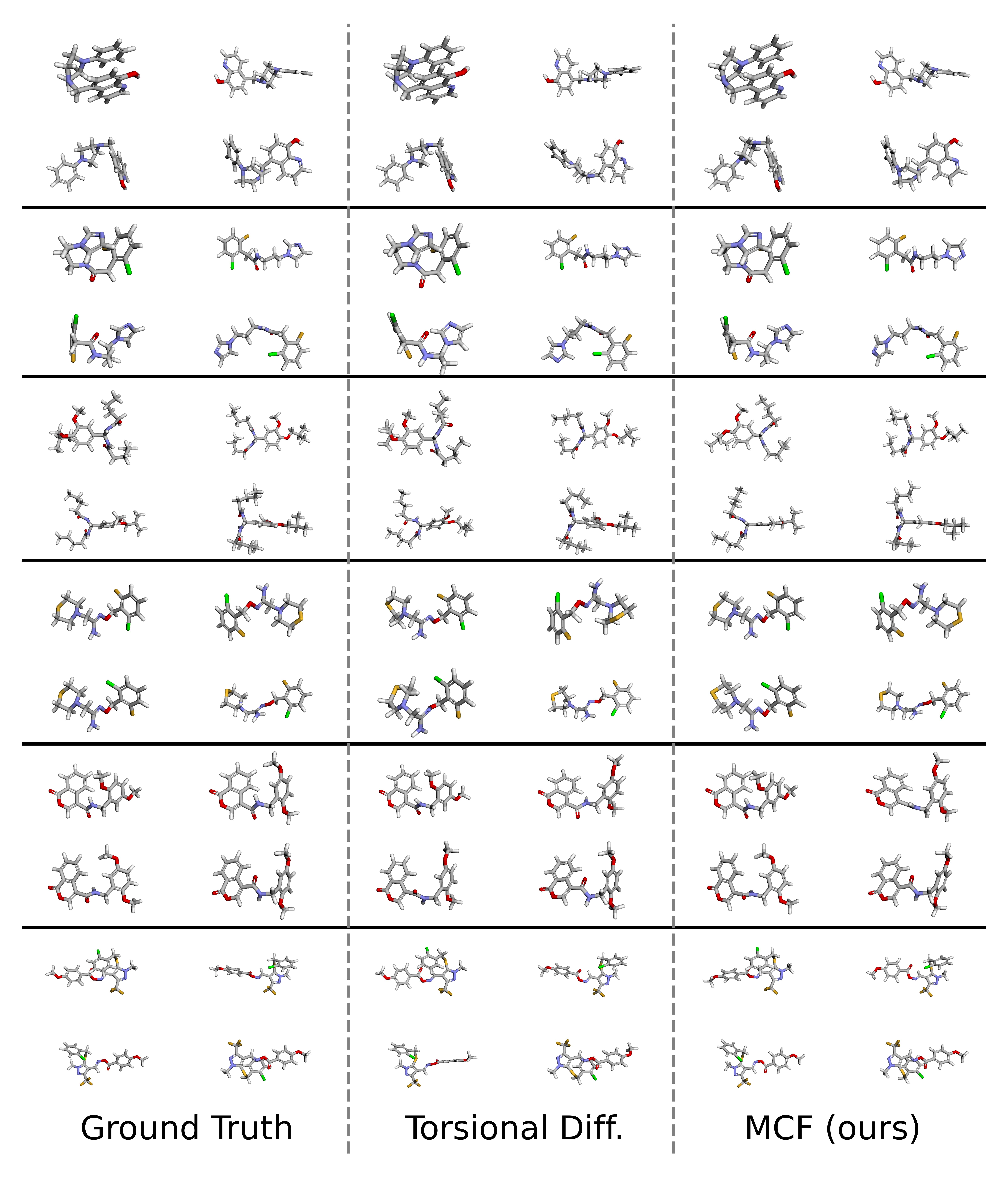}
    \caption{Examples of conformers of ground truth, Torsional Diff. samples, and {\model} samples from GEOM-DRUGS.}
    \label{fig:drugs_examples}
\end{figure*}

\begin{figure*}
    \centering
    \includegraphics[width=\textwidth]{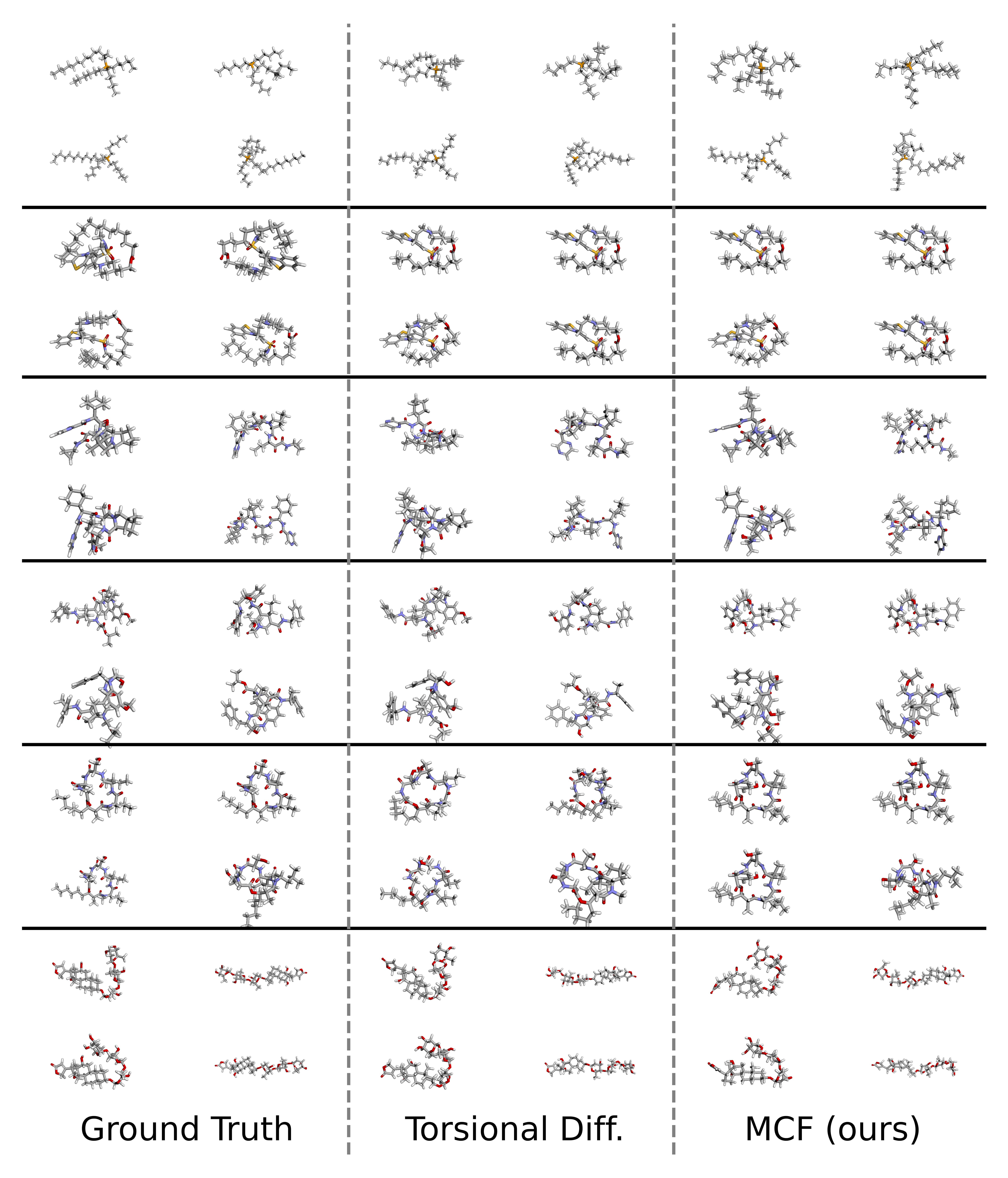}
    \caption{Examples of conformers of ground truth, Torsional Diff. samples, and {\model} samples from GEOM-XL.}
    \label{fig:xl_examples}
\end{figure*}

Fig.~\ref{fig:ddim_samples} show some examples of sampled conformers from {\model} with different sampling steps. It is illustrated that even with very limited sample steps, {\model} can still generate plausible conformers especially for the heavy atoms. We also show examples of conformers from ground truth, Torsional Diff., and our {\model}. Fig.~\ref{fig:drugs_examples} and \ref{fig:xl_examples} depict samples from GEOM-DRUGS and GEOM-XL, respectively. We found the samples that are most aligned with ground truth and plot them side by side. 

\end{document}

%% file: math_commands.tex
\usepackage{amsmath,amsfonts,bm}

\def\eqref#1{equation~\ref{#1}}

\def\1{\bm{1}}

\def\rmC{{\mathbf{C}}}

\def\rmI{{\mathbf{I}}}

\def\rmQ{{\mathbf{Q}}}

\def\rmY{{\mathbf{Y}}}

\def\vzero{{\bm{0}}}

\def\vy{{\bm{y}}}
\def\vz{{\bm{z}}}

\DeclareMathAlphabet{\mathsfit}{\encodingdefault}{\sfdefault}{m}{sl}
\SetMathAlphabet{\mathsfit}{bold}{\encodingdefault}{\sfdefault}{bx}{n}

